\documentclass[11pt,a4paper]{article}
\usepackage{jheppub}

\usepackage{psfrag}
\usepackage{subfigure}
\usepackage{picins}
\usepackage{bm}
\usepackage{dsfont} 
\usepackage{upref}
\usepackage{axodraw4j}
\usepackage{pstricks}
\usepackage{slashed}

\newcommand{\ddx}{\textrm{d}^d x\:}
\newcommand{\Tr}{\text{Tr}\:}
\newcommand{\tr}{\text{tr}\:}
\newcommand{\be}{\begin{equation}}
\newcommand{\ee}{\end{equation}}
\newcommand{\ba}{\begin{align}}
\newcommand{\ea}{\begin{align}}
\newcommand{\p}{\partial}
\newcommand{\sg}{\sqrt{g}}
\newcommand{\sgb}{\sqrt{\bar{g}}}
\newcommand{\gb}{\bar{g}}
\newcommand{\mn}{{\mu\nu}}
\newcommand{\rs}{{\rho\sigma}}
\newcommand{\Rk}{\mathcal{R}_k}
\newcommand{\ord}{\mathcal{O}}
\newcommand{\ZFk}{Z_{F,k}}
\newcommand{\sD}{\slashed{D}}
\newcommand{\id}{\mathds{1}}
\newcommand{\DS}{\mathcal{D}_\text{S}}
\newcommand{\RN}{R^{(0)}}
\newcommand{\ns}{n_S}
\newcommand{\mU}{\mathbf{U}} 
\newcommand{\bp}{\bm{\varphi}}
\newcommand{\cA}{\mathcal{A}}

\title{	On the physical mechanism underlying\\ Asymptotic Safety}

\author{Andreas Nink}
\author{and Martin Reuter}

\affiliation{PRISMA Cluster of Excellence \& Institute of Physics (THEP),\\
	University of Mainz, Staudingerweg 7, D-55099 Mainz, Germany}

\emailAdd{nink@thep.physik.uni-mainz.de}

\abstract{We identify a simple physical mechanism which is at the heart of Asymptotic
	Safety in Quantum Einstein Gravity (QEG) according to all available effective average
	action-based investigations. Upon linearization the gravitational field equations give
	rise to an inverse propagator for metric fluctuations comprising two pieces: a
	covariant Laplacian and a curvature dependent potential term. By analogy with
	elementary magnetic systems they lead to, respectively, dia- and paramagnetic-type
	interactions of the metric fluctuations with the background gravitational field. We
	show that above 3 spacetime dimensions the gravitational antiscreening occurring in QEG
	is entirely due to a strong dominance of the ultralocal paramagnetic interactions over
	the diamagnetic ones that favor screening. (Below 3 dimensions both the dia- and
	paramagnetic effects support antiscreening.) The spacetimes of QEG are interpreted as
	a polarizable medium with a ``paramagnetic'' response to external perturbations, and
	similarities with the vacuum state of Yang-Mills theory are pointed out. As a
	by-product, we resolve a longstanding puzzle concerning the beta function of Newton's
	constant in $2+\epsilon$ dimensional gravity.}


\preprint{MZ-TH/12-30}

\arxivnumber{1208.0031}

\begin{document}

\maketitle

\section{Introduction}
\label{sec:Intro}
While finding a consistent and predictive quantum theory of gravity is considered one
of the major challenges of today's theoretical physics it seems clear that still
substantial efforts are necessary in order to reach this goal \cite{Kiefer,Hamber-book}.
Presently a variety of approaches is explored, for instance loop quantum gravity and
spin foam models based on Ashtekar's variables \cite{A,R,T} or statistical mechanics
models based on causal dynamical triangulations \cite{CDT} or Regge calculus \cite{Regge}.
All of these approaches come with their specific advantages and drawbacks.
A scenario which is particularly attractive from the physics point of view is the idea
of Asymptotic Safety \cite{Weinberg} since it does not rely on any unproven assumptions
such as, say, higher dimensions, supersymmetry, or the existence of extended objects
such as strings or branes, for instance. However, its ultimate success will crucially
depend on whether we are able to understand the \emph{nonperturbative} dynamics of a
background independent quantum field theory sufficiently well.

In its form based upon the gravitational average action \cite{mr} the first step in the
Asymptotic Safety program consists in defining a coarse graining flow on an appropriate
theory space which comprises action functionals depending on the metric or similar field
variables. Then one searches for nontrivial fixed points of this flow by means of
functional renormalization group (RG) techniques.
If there is none, the idea fails right from the start. On the other hand, if such fixed
points exist, one must embark on the second step and try to construct a \emph{complete}
RG trajectory entirely within the theory space of well defined actions whereby the limit
corresponding to an infinite ultraviolet (UV) cutoff is taken at the fixed point in
question. In the successful case this trajectory defines a (candidate for a)
nonperturbatively renormalized quantum field theory whose properties and predictions
can be explored then. Furthermore, as the last step of the program one can use the
RG trajectory in order to construct a representation of the quantum theory in terms of
a UV-regularized functional integral; only then one will know the underlying Hamiltonian
system which, implicitly, got quantized by taking the UV limit at the fixed point
computed \cite{elisa1}.

During the past decade a large number of detailed studies of the gravitational RG flow
has been performed and significant evidence for the viability of the Asymptotic Safety
program was found. In particular, all investigations carried out to date unanimously
agree on the existence of a non-Gaussian fixed point (NGFP) at which the infinite cutoff
limit can be taken \cite{mr,souma,frank1,oliver,NJP,livrev,reviews,CPR}.

However, exciting and encouraging as they are, these findings are still somewhat
mysterious since there is no general physical or mathematical understanding yet as to
\emph{why} this fixed point should exist, rendering Quantum Einstein Gravity (QEG)
nonperturbatively renormalizable. In fact, most of the existing investigations pick a
certain truncated theory space and then calculate the $\beta$-functions describing the RG
flow on it. Typically this step is technically extremely involved; often it requires
developing new and non-standard computational tools and in any case it tends to be of
frightening algebraic complexity. This is particularly true for the bimetric truncations
\cite{elisa2,MRS1,MRS2,boundary} which need to be considered as a consequence of the
``paradoxical'' implementation of background independence by means of background fields
\cite{dewitt-books}. After the $\beta$-functions are found it is usually comparatively
easy to solve the differential equations they give rise to, and to study the resulting
RG flow. Only at this final point an interpretation in physical terms can be attempted.
The input, a specific truncation ansatz, is typically selected because it is ``natural''
in mathematical terms, for instance as part of a derivative expansion or an asymptotic
heat kernel series. Then, in order to systematically improve the ansatz, and to get an
idea of the physics encapsulated in it, it is necessary to re-compute the
$\beta$-functions within a modified truncation.

In this situation it would be very desirable to gain some intuitive understanding about
the features of a truncated action functional which are essential for exploring Asymptotic
Safety and which are not.

In this paper we take a first step in this direction by identifying a simple physical
mechanism which, according to all average action-based studies of Asymptotic Safety,
seems to underlie the formation of the crucial non-Gaussian RG fixed point in QEG.
We shall demonstrate that it owes its existence to a \emph{predominantly paramagnetic
interaction of the metric fluctuations with an external gravitational field.}

Let us explain the meaning of ``paramagnetic'' in this context. It is well known that
nonrelativistic electrons in an external magnetic field are described by the Pauli
Hamiltonian
\be
H_\text{P} = \frac{1}{2m}\,(\mathbf{p}-e\mathbf{A})^2 + \mu_\text{B}\,
	\mathbf{B}\cdot \bm{\sigma}\,.
\label{eqn:PauliHam}
\ee
It is equally well known that the first term on the RHS of (\ref{eqn:PauliHam}),
essentially the gauge covariant Laplacian, gives rise to the Landau diamagnetism of a
free electron gas, while the second term is the origin of the Pauli (spin) paramagnetism.
The former is due to the electrons' orbital motion, the latter to their spin alignment;
they are characterized by a negative ($\chi_\text{Landau-dia}<0$) and a positive
($\chi_\text{Pauli-para}>0$) magnetic susceptibility, respectively. An important result
is the relation between these two quantities,
\be
\chi_\text{Landau-dia} = -\frac{1}{3}\, \chi_\text{Pauli-para}\,,
\label{eqn:DiaVsPara}
\ee
implying that it is always the \emph{paramagnetic} component which ``wins'' and
determines the overall sign of the total susceptibility:
$\chi_\text{mag}\equiv \chi_\text{Landau-dia}+\chi_\text{Pauli-para} >0$.

From a more general perspective we should think of the Pauli Hamiltonian in position
space as a \emph{nonminimal} matrix differential operator, consisting of a covariant
Laplacian, $\bm{D}^2=(\bm{\nabla}-ie\mathbf{A})^2$, plus a non-derivative term
$\propto \mathbf{B}\cdot\bm{\sigma}$ which involves the ``curvature'' $\mathbf{B}$
of the ``connection'' $\mathbf{A}$.

This pattern persists when we move on to the relativistic analog of (\ref{eqn:PauliHam}),
the square of the Dirac operator
$\sD\equiv \gamma^\mu D_\mu \equiv \gamma^\mu(\p_\mu-ieA_\mu)$, namely
\be
\sD^2 = D^2-\frac{i}{2}\,e\,\gamma^\mu \gamma^\nu F_{\mu\nu}\,.
\label{eqn:QEDoperator}
\ee
This differential operator, too, is nonminimal. It comprises a covariant Laplacian,
$D^2\equiv D^\mu D_\mu$, responsible for the orbital motion-related ``diamagnetic''
effects, and a non-de\-ri\-va\-tive term which is non-diagonal in spinor space and causes
the ``paramagnetic'' effects.\footnote{From the mathematical point of view the operator
(\ref{eqn:QEDoperator}) generalizes to a large class of nonminimal second order operators
with similar properties, the Laplacians admitting a Weitzenb\"{o}ck decomposition
\cite{Lawson}. They play an important r\^{o}le in differential geometry
(Lichnerowicz- and Bochner theorems).}

The same physics based distinction of orbital motion vs{.} spin alignment effects can
also be made for bosonic systems. Let us consider an {\sf SU}($N$) gauge field
$A_\mu^a \,$, $a=1,\cdots,N^2-1$, governed by the classical Yang-Mills Lagrangian
$\propto (F_{\mu\nu}^a)^2$. If we expand $A_\mu^a$ at some background field
$\bar{A}_\mu^a$, small fluctuations about this background, $\delta A_\mu^a$, are described
by a quadratic action of the form $\int \delta A_\mu^a(\cdots)\delta A_\nu^b$. In Feynman
gauge the kernel $(\cdots)$ is given by the nonminimal operator
\be
\left(\bar{D}^2\right)^{ab}\delta_\mu^{~\nu} - 2ig\bar{F}^{ab~\nu}_{~~\mu}\,.
\label{eqn:YMoperator}
\ee
Here $\bar{D}^2\equiv \bar{D}^\mu \bar{D}_\mu$ and $\bar{F}^{ab~\nu}_{~~\mu}$ are built
from the background field. Clearly the operator (\ref{eqn:YMoperator}) has the same
structure as (\ref{eqn:QEDoperator}), namely a Laplacian, covariantized by means of a
certain connection, plus a multiplicative term linear in the corresponding curvature.

The fluctuations $\delta A_\mu^a$ have two qualitatively different interactions with
the background field $\bar{A}_\mu^a$\;: an orbital one
$\propto \int\delta A \, \bar{D}^2 \delta A$ related to their spacetime dependence, and
an ultralocal one $\propto\int \delta A \, \bar{F} \delta A$ which is sensitive to the
orientation of the fluctuations relative to the background in color space. In this sense,
one considers the vacuum a color magnetic medium where the corresponding susceptibility
turns out to be proportional to the $\beta$-function of the Yang-Mills coupling. The
occurrence of two different types of contributions in (\ref{eqn:YMoperator}) raises the
question whether in Yang-Mills theory, too, there are specific effects which can be
attributed to the first, ``diamagnetic'', and the second, ``paramagnetic'', term
separately \cite{Nielsen}.

A well known example where this can be done
\cite{polyakov-book,Johnson,huang-book,Gottfried-Weisskopf,dittrich} is asymptotic
freedom \cite{PolitzerGrossWilczek}. In fact, the one-loop Yang-Mills $\beta$-function
can be presented in the decomposed form
\be
\beta_{g^2} = - \frac{N}{24\pi^2}\,\big[\,12-2+1\,\big]\,g^4 \,,
\ee
where the ``+12'' is due to the fluctuations' paramagnetic interaction, the ``-2'' stems
from the diamagnetic one, and the ``+1'' comes from the Faddeev-Popov ghosts. The para-
and diamagnetic contributions come with an opposite sign, but since the former are six
times bigger than the latter, it is the paramagnetic interaction that determines the
overall negative sign of $\beta_{g^2}$. In this respect $\beta_{g^2}$ is analogous to the
magnetic susceptibility $\chi_\text{mag}$ whose sign is also determined by the
competition of para- and diamagnetic effects, the clear winner being paramagnetism.

Thus we can say that in Yang-Mills theory \emph{asymptotic freedom is due to the
predominantly paramagnetic interaction of gauge field fluctuations with external fields.}

\medskip
In this paper we are going to show that the Asymptotic Safety of QEG is, in the sense of
this magnetic analogy, very similar to the asymptotic freedom of Yang-Mills theory, the
main difference being that the Gaussian fixed point implicit in perturbative
renormalization is replaced by a nontrivial one now. The similarities are most clearly
seen in the Einstein-Hilbert truncation of the QEG theory space \cite{mr}. The dynamics
of fluctuations $h_\mn(x)$ about a prescribed metric background $\bar{g}_\mn(x)$ is given
by a quadratic action $\propto \int\sgb\; h_\mn(\cdots)\,h^{\rho\sigma}$ whose kernel
$(\cdots)$ is found by expanding the Einstein-Hilbert action to second order. The result
with a harmonic gauge fixing is again a nonminimal matrix differential operator with a
clear separation of ``dia-'' vs. ``paramagnetic'' couplings to the background:
\be
-\bar{K}^\mn_{~~\rho\sigma} \, \bar{D}^2 + \bar{U}^\mn_{~~\rs}\,.
\label{eqn:GravOperator}
\ee
Here $\bar{D}^2\equiv\gb^\mn \bar{D}_\mu\bar{D}_\nu$, where $\bar{D}_\mu$ is the
covariant derivative with respect to the Levi-Civita connection given by $\gb_\mn$,
and $\bar{U}^\mn_{~~\rs}$ is a tensor built from the background's curvature tensor,
\be
\begin{split}
\bar{U}^\mn_{~~\rs} = \frac{1}{4}\left[\delta_\rho^\mu \delta_\sigma^\nu +
	\delta_\sigma^\mu \delta_\rho^\nu - \gb^\mn \gb_\rs \right] \left(\bar{R}-2\Lambda_k\right)
	+ \frac{1}{2}\left[\gb^\mn \bar{R}_\rs + \gb_\rs \bar{R}^\mn\right]& \\
	- \frac{1}{4}\left[\delta_\rho^\mu\bar{R}^\nu_{~\sigma} + \delta_\sigma^\mu\bar{R}^\nu_{~\rho}
	+ \delta_\rho^\nu\bar{R}^\mu_{~\sigma} + \delta_\sigma^\nu\bar{R}^\mu_{~\rho}\right]
	- \frac{1}{2}\left[\bar{R}^{\nu~\mu}_{~\rho~\sigma} + \bar{R}^{\nu~\mu}_{~\sigma~\rho}\right]&.
\end{split}
\label{eqn:UGrav}
\ee
Furthermore, $\bar{K}^\mn_{~~\rs} = \frac{1}{4}\left[\delta_\rho^\mu \delta_\sigma^\nu
+\delta_\sigma^\mu \delta_\rho^\nu - \gb^\mn \gb_\rs \right]$.
The r\^{o}le of $\beta_{g^2}$ in 4D Yang-Mills theory is played by the anomalous dimension
$\eta_N$ of Newton's constant now. Here, too, it is possible to disentangle dia- and
paramagnetic contributions. Again they come with opposite signs, and the paramagnetic
effects turn out much stronger than their diamagnetic competitors or the ghosts.

As a consequence, the negative sign of $\eta_N$ governing the RG running of Newton's
constant, crucial for Asymptotic Safety and gravitational antiscreening, originates from
the paramagnetic interaction of the metric fluctuations with their background (or
``condensate''). The diamagnetic effects counteract the antiscreening trend and the
formation of an NGFP, but they are too weak to overwhelm the paramagnetic ones. This is
what we shall call \emph{paramagnetic dominance.}

Thus we see that it is the nonminimal part of the fluctuations' inverse propagator
(\ref{eqn:GravOperator}) that determines the essential features of nonperturbative
gravity. This mechanism becomes manifest in the special case of three dimensions.
There are no propagating degrees of freedom in $d=3$, that is no gravitational waves.
Nevertheless the couplings parametrizing a generic effective average action, such as the
scale dependent Newton constant, for instance, show a nontrivial RG running, very much
like in $4$ dimensions. In this paper we shall resolve this apparent paradox as follows.
Diagrammatically speaking, the non-existence of propagating physical gravitons is a result
of the antagonistic effects of the metric fluctuations and the ghosts circulating
inside loops. If, and only if, $d=3$ the contributions due to their respective $\bar{D}^2$
kinetic terms cancel precisely. Hence the net ``diamagnetic'' contribution vanishes
exactly. Allowing for a non-flat background, however, the fluctuation modes couple
nontrivially to its curvature. As a result the entire RG behavior is determined by this
``paramagnetic'' interaction.

We shall also address the special case of $2+\epsilon$ dimensions in detail since in the
past there has been a certain confusion about the correct leading order coefficient in
the $\beta$-function for Newton's constant. This longstanding controversy will be
resolved by attributing the different results to a different treatment of the
paramagnetic interaction term.

To complete the picture of ``paramagnetic dominance'' we shall then leave the old-type
single-metric computations and investigate an example of a bimetric truncation ansatz
\cite{elisa2,MRS1,MRS2}. For this purpose we consider an action
$\propto \int \sg\; A \left( -D^2 + \xi R \right)A$,
describing scalar fields $A$, nonminimally coupled to the metric, and carefully
distinguish the dynamical from the background metric during the calculation. The dynamics
of $A$ is determined by the nonminimal differential operator
\be
-D^2+\xi R\,,
\ee
similar to the ones mentioned above. Varying the parameter $\xi$ which is treated as a
scale independent constant we shall study the impact of the ``paramagnetic'' term on the
resulting RG flow. As we will demonstrate, only for $\xi$ large enough one finds a fixed
point with a positive value of Newton's constant.
\medskip

Before presenting the details of the mechanism mentioned above we shall briefly summarize
some essentials of the calculational method we employ.

In our approach the scale dependence due to the renormalization or ``coarse graining''
processes is studied by means of the effective average action $\Gamma_k$
\cite{avact}.\footnote{Unless stated otherwise, all bare and effective actions considered
in this paper refer to a Euclidean spacetime.} The basic feature of this action
functional consists in integrating out all quantum fluctuations of the underlying fields
from the UV down to an infrared cutoff scale $k$. In a sense, $\Gamma_k$ can be
considered the generating functional of the 1PI correlation functions that take into
account the fluctuations of all scales larger than $k$. Consequently, for $k=0$ it
coincides with the usual effective action, $\Gamma_{k=0}=\Gamma$. On the other hand, in
the limit $k\rightarrow\infty$, $\Gamma_k$ approaches the bare action $S$, apart from a
simple, explicitly known correction term \cite{elisa1}.

Starting from the functional integral definition of $\Gamma_k$ one may investigate its
scale dependence by taking a $k$-derivative, which results in the exact functional
renormalization group equation (FRGE) \cite{avact,YM-EAA,YM-AA}
\be
\p_t \Gamma_k = \frac{1}{2}\, \text{STr}\left[ \left( \Gamma_k^{(2)}+ \Rk \right)^{-1}
	\p_t \Rk \right] \,.
\label{eqn:FRGE}
\ee
Here we introduced the RG time $t=\ln k$. The differential operator $\Rk$ in
(\ref{eqn:FRGE}) comprises the infrared cutoff: in the corresponding path integral the
bare action $S$ is replaced by $S+\Delta_k S$ where the cutoff action added is quadratic
in the fluctuations $\phi$, $\Delta_k S \propto \int \phi\Rk\phi$. Furthermore,
$\Gamma_k^{(2)}$ denotes the ``matrix'' of second functional derivatives with respect to
the dynamical fields. The functional supertrace in (\ref{eqn:FRGE}) includes a trace over
all internal indices, and a sum over all dynamical fields, with an extra minus sign for
the Grassmann-odd ones.

The effective average action ``lives'' in the infinite dimensional ``theory space'' of
all action functionals depending on a given set of fields and respecting the required
symmetries. Its RG evolution determined through eq.\ (\ref{eqn:FRGE}) amounts to a curve
$k \mapsto \Gamma_k$ in theory space. Since the FRGE leads in general to a system of
infinitely many coupled differential equations, one usually has to resort to truncations
of the theory space. For this purpose $\Gamma_k$ is expanded in a basis of field
monomials $P_\alpha[\,\cdot\,]$, i.e.
$\Gamma_k[\,\cdot\,]=\sum_\alpha c_\alpha(k) P_\alpha[\,\cdot\,]$,
but then one truncates the sum after a finite number of terms. Thus the scale dependence
of $\Gamma_k$ is described by finitely many running couplings $c_\alpha(k)$. Projecting
the RHS of (\ref{eqn:FRGE}) onto the chosen subspace of theory space the resulting system
of differential equations for the couplings remains finite, too. In this way we obtain
the $\beta$-functions for the couplings $c_\alpha$.

In most studies of quantum gravity carried out so far the monomials $P_\alpha[\,\cdot\,]$
retained in the truncation ansatz were local functionals of the metric, typically powers
of the curvature tensor and its contractions. By analogy with Yang-Mills theory this
should be a valid approximation at high scales $k$ when the trajectory is close to its
UV fixed point. In fact, for $k\rightarrow\infty$ the average action approaches
$\Gamma_*$ which is essentially the same as the bare action $S$, that, at least in a
standard field theory such as QCD, is a simple local functional. On the other hand, when
$k$ approaches the QCD (confinement) scale $\Lambda_\text{QCD}$ \emph{from above} purely
local truncations become insufficient, and finally, for $k\rightarrow 0$ at the latest,
all the nonlocal terms are generated which must be present in the ordinary 1PI generating
functional $\Gamma_{k=0}\equiv\Gamma$. In QEG the situation is believed to be similar,
with the Planck mass playing a r\^{o}le analogous to $\Lambda_\text{QCD}$. (See
\cite{frank1} for a discussion.) Since we are mostly interested in UV aspects of the RG
flow here, nonlocal terms are unlikely to be qualitatively important.

In order to illustrate the main idea of this work we shall thus use simple local
truncation ans\"{a}tze for the respective system, for instance the Yang-Mills action
$\propto \int (F_\mn^a)^2$ for an {\sf SU}$(N)$ gauge theory, or the Einstein-Hilbert
action in the case of gravity, each one furnished with running couplings.

\medskip
The remaining sections in this article are organized as follows. Section \ref{sec:QED-YM}
demonstrates the idea of distinguishing between the two ``magnetic'' contributions with
the help of two examples: fermions in QED and gauge bosons in Yang-Mills theory. In
section \ref{sec:QEG} we perform the analogous analysis for gravity and focus on the
question which terms render QEG, in the Einstein-Hilbert truncation, asymptotically safe.
Section \ref{sec:Bimetric} is devoted to the investigation of a matter induced bimetric
action. In section \ref{sec:QEGvacuum} we interpret the spacetimes of QEG as a
polarizable medium, emphasizing certain analogies with Yang-Mills theory. Our main
results are summarized in section \ref{sec:conclusion}.

\section{Paramagnetic dominance: known examples}
\label{sec:QED-YM}
This section is meant to demonstrate our method by means of two well known examples: QED
and Yang-Mills theory.
 
By calculating the interaction energy between two (generalized) charges it is possible
to define analogs of the electric and magnetic susceptibility also for other field
theories than electrodynamics, for instance Yang-Mills theory \cite{Gottfried-Weisskopf}.
From a renormalization point of view this can be used to establish a connection between
the susceptibility and the $\beta$-function. Let us consider a massless charged field
with spin $S$ and renormalized charge $g$. The lowest order of the $\beta$-function for
$g^2$ is quartic in $g$, so that one can expand $\beta_{g^2}=\beta_0 g^4+\ord(g^6)$.
Then one finds a relation for the magnetic susceptibility,
$\chi_\text{mag}\propto \beta_0$, where $\beta_0$ is given by \cite{Nielsen}
\be
\beta_0 = - \frac{(-1)^{2S}}{4\pi^2} \,\left[ (2S)^2-\frac{1}{3}\right] \,.
\label{eqn:NielsenHughes}
\ee
Here the first term, $(2S)^2$, is due to the ``paramagnetic'' interaction, while the
$-\frac{1}{3}$ is the ``diamagnetic'' contribution.
For spin-$\frac{1}{2}$ fermions eq.\ (\ref{eqn:NielsenHughes}) reduces to
$\beta_0^\text{QED}=\frac{1}{4\pi^2}\left[1-\frac{1}{3}\right]$, reproducing relation
(\ref{eqn:DiaVsPara}): $\beta_0^\text{dia}=-\frac{1}{3}\beta_0^\text{para}$.
A similar result is obtained in QCD. One has to determine the weighted sum over all
charged gluons contributing to (\ref{eqn:NielsenHughes}) \cite{huang-book}. Then the
gluon-only part of the $\beta$-function at lowest order is seen to assume the form
$\beta_0^\text{QCD}=-\frac{1}{8\pi^2}\big[12-1\big]$.

In the following we shall rederive these findings within the FRGE approach, and present
the analysis in a way which lends itself to a generalization to gravity.

\subsection{Paramagnetic dominance in QED}
\label{sec:QED}
In order to derive the QED $\beta$-function we have to choose an appropriate truncation
for the effective average action $\Gamma_k$. Since we are interested only in the lowest
order of the $\beta$-function, it is sufficient to consider the simple ansatz, in 4
Euclidean dimensions,
\be
\Gamma_k[A,\bar{\psi},\psi] = \int \textrm{d}^4 x\,\left[ \ZFk\,\frac{1}{4}\, F_\mn F^\mn
	+\bar{\psi}\, \sD \,\psi \right]\,.
\label{eqn:GammaQED}
\ee
Here $F_\mn\equiv \p_\mu A_\nu - \p_\nu A_\mu$ is the field strength tensor built from
the gauge field $A_\mu$, $\psi$ denotes the fermion field, and $\ZFk$ is a wave function
renormalization constant. We do not include any mass term since it would not change our
conclusions qualitatively. Using (\ref{eqn:GammaQED}) we determine the influence of the
fermion field on the running of $\ZFk$, that is, on the propagation of the gauge field.

With the above truncation it is sufficient to quantize the fermions, keeping the gauge
field as a classical background. As a result the FRGE (\ref{eqn:FRGE}) can be written in
the form
\be
\p_t \Gamma_k = - \text{Tr}\left[ \left( \Gamma_k^{(2)}+ \Rk \right)^{-1}_{\bar{\psi}\psi}
	\, \p_t \Rk \right]\,,
\label{eqn:FRGEQED}
\ee
where the trace is over fermionic fluctuation modes only, and the second functional
derivative is given by $\left(\Gamma_k^{(2)}\right)_{\bar{\psi}\psi}(x,y)\equiv
\delta/\delta\psi(x) \big[\delta/\delta\bar{\psi}(y)\,\Gamma_k\big]$ with suppressed
spinor indices. It is convenient to reexpress the RHS of (\ref{eqn:FRGEQED}) such
that $\Gamma_k^{(2)}=\sD$ makes its appearance via its square only. For this purpose one
can exploit the formal operator trace identity $\Tr\ln(a\sD+b\id)=\frac{1}{2}\,
\Tr\ln(-a^2\sD{}^2+b^2\id)$, which is valid for an adequate regularization of the trace
\cite{dittrich,mrtheta}. Provided that $\Rk \propto \id$ we obtain
\be
\Tr\left( \frac{\p_t\Rk}{\Gamma_k^{(2)}+\Rk} \right) = \frac{1}{2}\, \Tr
	\left( \frac{\p_t(\Rk^2)}{-\sD^2+\Rk^2} \right)\,,
\label{eqn:TrRevised}
\ee
with the cutoff operator $\Rk \equiv \Rk(-\sD{}^2) \equiv k \, \RN(-\sD{}^2/k^2)$. Here
$\RN$ is an arbitrary cutoff shape function interpolating smoothly between $\RN(0)=1$ and
$\RN(\infty)=0$.

At this point we see how the separation of the magnetic contributions arises naturally:
the operator $\sD{}^2$ appearing under the functional trace of (\ref{eqn:TrRevised})
satisfies the relation
\be
\sD^2=D^2\id - \frac{i}{2}\, \bar{e}\, \gamma^\mu \gamma^\nu F_\mn \,,
\label{eqn:DSlashed2}
\ee
where $\bar{e}$ denotes the bare charge. The first, minimal, term in
(\ref{eqn:DSlashed2}) is referred to as the diamagnetic part and the second, nonminimal,
one as the paramagnetic part. We shall now disentangle these different contributions up
to the final result for the $\beta$-function.

Applying standard heat kernel techniques \cite{heat-kernel} we can project
the trace occurring in (\ref{eqn:FRGEQED}) and (\ref{eqn:TrRevised}) onto a basis of
monomials constructed from the field strength tensor. A comparison of the coefficients
of $\int \textrm{d}^4 x\, F_\mn F^\mn$ in (\ref{eqn:FRGEQED}) yields a relation for the
anomalous dimension $\eta_F \equiv -\p_t \ln \ZFk\,$. If we define the renormalized
charge by $e^2=\ZFk^{-1}\,\bar{e}^2$, we finally arrive at
\be
\p_t e^2 = \beta_{e^2} = \frac{1}{4\pi^2}\, \Bigg[ \bigg\{1\bigg\}_\text{para} +
	\bigg\{-\frac{1}{3}\bigg\}_\text{dia} \;\Bigg]\, e^4 \,.
\label{eqn:betaQED}
\ee
We use curly brackets in (\ref{eqn:betaQED}) in order to separate and label the different
contributions to the total sum. This notation will be employed in the following sections,
too. As we expected, the result according to (\ref{eqn:NielsenHughes}) is retrieved.
In particular, we clearly see the relation
$\beta_{e^2}^\text{dia} = -\frac{1}{3}\,\beta_{e^2}^\text{para}$.
The positive sign of $\beta_{e^2}$ is a crucial feature of QED. It is responsible for
screening effects, and for a possible singularity of the renormalized charge emerging at
a large but finite energy scale, the Landau pole \cite{GiesJaeckel}.
Since it is the paramagnetic term in (\ref{eqn:betaQED}) that dictates the overall sign,
we can conclude that the qualitative properties of QED, particularly its asymptotic
behavior, are determined by paramagnetism. We emphasize that these findings, based on
(\ref{eqn:betaQED}), are \emph{universal}, i.e.\ they are independent of the cutoff shape
function $\RN$ we choose.

\subsection{Paramagnetic dominance in Yang-Mills theory}
\label{sec:YM}
Now we transfer the concepts employed above for QED to the non-Abelian case, and
investigate in particular the origin of asymptotic freedom in Yang-Mills theory, viewing
its vacuum as a color magnetic medium. We keep the spacetime dimension $d$ arbitrary.
For $d\neq 4$ the Yang-Mills coupling is dimensionful and so the $\beta$-function of its
dimensionless counterpart contains the classical scaling dimension $d-4$ besides the
anomalous dimension $\eta_F$:
\be
\p_t g^2 = \beta_{g^2} \equiv (d-4+\eta_F)g^2 \,.
\ee
Since it is $\eta_F$ that comprises the quantum effects we are interested in, we shall
discuss the different ``magnetic'' contributions at the level of $\eta_F$ rather than the
$\beta$-function.

\piccaption{Schematic diagram of the coupling of YM-field fluctuations to the background.
\label{fig:YM-Feyn}}
\pichskip{2em}
\parpic[l]{
\fcolorbox{white}{white}{
	\begin{picture}(136,60) (100,-90)
    \SetWidth{0.8}
    \SetColor{Black}
    \Gluon(111,-41)(159,-41){5}{4}
    \SetWidth{1.0}
    \Vertex(159,-40){2.236}  
    \Text(164,-92)[lb]{\small{\Black{$\bar{A}_\mu$}}}
    \Line[dash,dashsize=2](159,-89)(159,-43)
    \Text(113,-58)[lb]{\small{\Black{$a_\mu$}}}
    \Text(199,-58)[lb]{\small{\Black{$a_\mu$}}}
    \SetWidth{0.8}
    \Gluon(160,-41)(208,-41){5}{4}
  \end{picture}
}}
Within the nonperturbative setting the calculation proceeds as follows. If we employ the
background formalism, i.e.\ split the dynamical gauge field $A_\mu^a$ into a sum of a
rigid background $\bar{A}_\mu^a$ and a fluctuation $a_\mu \equiv \delta A_\mu^a$, the
propagation of the fluctuating field is crucially influenced by its interaction with the
background. In this regard the background assumes the r\^{o}le of an external color
magnetic field that couples to the fluctuations and probes their properties, see figure
\ref{fig:YM-Feyn}. Our goal is to determine the scale dependence of the corresponding
coupling constant.

Choosing the gauge group {\sf SU}$(N)$ we construct gauge invariant combinations of the
gauge field $A_\mu^a$ as candidates for appropriate action monomials. Here it turns out
sufficient to follow ref.\ \cite{YM-EAA} and consider a simple truncation for $\Gamma_k$
which consists of the usual Yang-Mills action, equipped with a scale dependent prefactor,
plus a gauge fixing term:
\be
\Gamma_k[A,\bar{A}\,] = \int \ddx \left\{ \frac{1}{4}\, \ZFk\, F_\mn^a[A] F_a^\mn[A]
	+ \frac{\ZFk}{2\alpha_k} \left[ D_\mu[\bar{A}](A^\mu-\bar{A}^\mu) \right]^2 \right\} \,.
\label{eqn:GammaYM}
\ee
Here the field strength tensor is given by
$F_\mn^a[A] = \p_\mu A_\nu^a - \p_\nu A_\mu^a + \bar{g}\,f_{bc}{}^a\, A_\mu^b A_\nu^c$
with the bare charge $\bar{g}$ and the structure constants $f_{bc}{}^a$. The gauge fixing
parameter $\alpha_k$ will be set to the constant value $1$ in the following.

For truncations of the type (\ref{eqn:GammaYM}) the general, exact FRGE for Yang-Mills
fields \cite{YM-EAA} boils down to the following decomposed form which treats gauge boson
and ghost contributions separately:
\be
\begin{split}
\p_t\Gamma_k[A,\bar{A}\,] = &\;\frac{1}{2}\, \Tr\left\{ \left( \Gamma_k^{(2)}[A,\bar{A}\,]
	+ \Rk[\bar{A}\,] \right)^{-1} \p_t \Rk[\bar{A}\,] \right\} \\
& - \Tr \left\{ \left( \DS[\bar{A}\,]	+ \Rk^\text{gh}[\bar{A}\,] \right)^{-1}
	\p_t \Rk^\text{gh}[\bar{A}\,] \right\} \,,
\end{split}
\label{eqn:FRGE-YM}
\ee
where $(\DS[\bar{A}\,])^a_{~b} \equiv -(D_\mu[\bar{A}\,]D^\mu[\bar{A}\,])^a_{~b}$.
Note the different arguments in the respective cutoff operators which are chosen as in
\cite{YM-EAA}. The gauge boson cutoff depends on
\be
\mathcal{D}_T[\bar{A}\,] \equiv -D^2[\bar{A}\,] + 2i \bar{g}\, F[\bar{A}\,] \,,
\label{eqn:DT}
\ee
a color matrix in the adjoint representation, while $\Rk^\text{gh}[\bar{A}\,]\equiv
\Rk\big(\DS[\bar{A}\,]\big)$ for the ghosts.

After taking the second functional derivative in (\ref{eqn:FRGE-YM}) we may identify
$\bar{A}=A$, project the traces onto the functional $\int\ddx F_\mn^a[A] F_a^\mn[A]$, and
deduce the running of $\ZFk$. With the gauge fields identified, $\Gamma_k^{(2)}$ reduces
to
\be
\Gamma_k^{(2)}[A] \equiv \frac{\delta^2}{\delta A^2} \Gamma_k[A,\bar{A}\,]
	\Big|_{\bar{A}=A} = \ZFk\,\mathcal{D}_T[A] = \ZFk \left( -D^2 + 2i\bar{g}\,F \,\right)
	\,.
\label{eqn:YMGammatwo}
\ee

We observe that the operator (\ref{eqn:YMGammatwo}) has a similar form as its QED analog
(\ref{eqn:DSlashed2}). Thus an obvious notion of ``dia-'' vs.\ ``paramagnetic''
contributions suggests itself: the first term of the RHS in (\ref{eqn:YMGammatwo})
represents diamagnetic interactions, and the second, nonminimal, term paramagnetic
ones. The only difference compared to the fermions in QED occurs due to the additional
ghost term in the FRGE. Since the ghost analog of (\ref{eqn:YMGammatwo}) is a minimal
operator, $\DS = -D^2$, its induced effects will be referred to as ``ghost-diamagnetic''
in the following.

Again we expand the traces in (\ref{eqn:FRGE-YM}) as a heat kernel series, compare the
coefficients of $\int\ddx F_\mn^a[A] F_a^\mn[A]$, and obtain a differential equation
describing the scale dependence of $\ZFk\,$.\footnote{See \cite{YM-EAA} for the details
of the calculation.} In terms of the dimensionless renormalized charge
$g^2 \equiv k^{d-4}\ZFk^{-1}\, \bar{g}^2$ this results in a relation for the anomalous
dimension $\eta_F \equiv -\p_t \ln \ZFk \,$,
\be
\eta_F(g) = - \frac{1}{3}\,(4\pi)^{-d/2}\, N \, \Phi_{d/2-2}^1(0) \Big[ \,
	\big\{24\big\}_\text{para} + \big\{-d\big\}_\text{dia}	+ \big\{2\big\}_\text{ghost-dia}
	\Big]	g^2 + \ord(g^4) \,.
\label{eqn:etaYM}
\ee
The factor $\Phi_{d/2-2}^1(0)$ in (\ref{eqn:etaYM}) denotes one of the standard threshold
functions, evaluated at vanishing argument. In general they are defined by \cite{mr}
\be
\Phi_n^p(w) \equiv \frac{1}{\Gamma(n)}\int_0^\infty dz\;
	z^{n-1}\frac{\RN(z)-z\RN{}'(z)}{\left[ z + \RN(z) + w \right]^p}\;,\quad n>0,
\label{eqn:Phi}
\ee
with the cutoff shape function $\RN$. For later use we also introduce
\be
\widetilde\Phi_n^p(w) \equiv \frac{1}{\Gamma(n)}\int_0^\infty dz\; z^{n-1}
	\frac{\RN(z)}{\left[ z + \RN(z) + w \right]^p} \;,\quad n>0.
\label{eqn:PhiTilde}
\ee
In the case $n=0$ we define $\Phi_0^p(w) = \widetilde\Phi_0^p(w) = (1+w)^{-p}$.
The threshold functions $\Phi_n^p$ and $\widetilde\Phi_n^p$ are \emph{positive} for all
$n$, in particular, concerning eq.\ (\ref{eqn:etaYM}), $\Phi_{d/2-2}^1(0)>0$ for any $d$.
In a generic dimension the numerical value of $\Phi_{d/2-2}^1(0)$ depends on the shape
function $\RN$. The case $d=4$ is special in that $\Phi^1_0(0)=1$ for any $\RN$.

We emphasize the importance of relation (\ref{eqn:etaYM}). For all $d$ smaller than $24$
we find the paramagnetic part to be dominant. With regard to relative signs, the
diamagnetic effect counteracts the paramagnetic and the ghost one. However, the
diamagnetic contribution is subdominant up to the ``critical'' dimension $d=26$ which has
$\eta_F = 0$. Hence, for $d < 26$ the anomalous dimension is negative, and this is
basically due to the paramagnetic term. In turn, it is this sign that determines the
qualitative behavior of the coupling $g$ at high energies. Therefore, one can say that
\emph{paramagnetism decides about whether or not the Yang-Mills theory is asymptotically
free/safe}.

The total \emph{diamagnetic} contribution in (\ref{eqn:etaYM}) is proportional to
$(d-2)$. This reflects the number of propagating (!) physical gauge bosons: for every
color direction, the $d$ originally available degrees of freedom of each gauge field are
reduced by 2 units if one exploits gauge invariance and the freedom to perform a residual
gauge transformation on shell \cite{weinberg-book}. This leads to $(d-2)$ degrees of
freedom, similar to the situation of a photon in electrodynamics. The vanishing number of
both propagating and physical degrees of freedom in $d=2$ corresponds to a vanishing
total diamagnetic contribution, such that $\eta_N$ is given entirely by the paramagnetic
term. We will encounter an analogous behavior for gravity in $d=3$ later on.

Finally, we focus on the case $d=4$. Then $\eta_F$ becomes universal
since $\Phi_0^1(0)=1$ is independent of the cutoff, and so we obtain
$\eta_F = - \frac{N}{24\pi^2}\,\big[\, \{12\}_\text{para} + \{-2\}_\text{dia}
+ \{1\}_\text{ghost-dia}\, \big]g^2 + \ord(g^4)$, or, equivalently,
\be
\beta_{g^2} = - \frac{N}{24\pi^2}\,\Big[ \,\big\{12\big\}_\text{para}
	+ \big\{-2\big\}_\text{dia}	+ \big\{1\big\}_\text{ghost-dia}	\Big]	g^4 + \ord(g^6) \,.
\label{eqn:betaYM}
\ee
The crucial overall minus sign driving $g$ to zero in the high energy limit results from
the first term of the sum. Thus we can conclude for four-dimensional Yang-Mills theory
that \emph{asymptotic freedom occurs only due to the paramagnetic interactions}.
In the QCD case with $N=3$ we see that (\ref{eqn:betaYM}) is in agreement with eq.\
(\ref{eqn:NielsenHughes}).

As for higher dimensions, the $F^2$-truncation used here leads to a non-Gaussian fixed
point $g_* \neq 0$. There $\beta_{g^2}$ vanishes by virtue of $d-4+\eta_F(g_*)=0$ for
$4<d<26$. According to an improved truncation \cite{GiesYM} this NGFP seems likely to
disappear for dimensionalities too far above $4$.

\medskip
Let us recapitulate. We considered the inverse propagator, an operator of the form
$-\bar{D}^2+U$ with a potential term $U$. It consists of two parts, a minimal one of
Laplace type and a nonminimal one. The effects induced by these different parts, in
particular their impact on the $\beta$-function, are called dia- and paramagnetic,
respectively. We found the latter to prevail in QED and Yang-Mills theory. Dia- and
paramagnetism, in this sense, correspond to rather different types of interactions the
quantized field fluctuations have with their classical background: via their spacetime
modulation, measured by $\bar{D}^2$ in the ``dia'' case, and by aligning their internal
degrees of freedom to the external field in the ``para'' case.

Before identifying the same mechanism also in gravity we shall mention some subtleties
concerning the notion of dia- and paramagnetic media.

\subsection{The vacuum as a magnetic medium}
\label{sec:vacuum}
The results of the previous subsections may be seen as a confirmation of eq.\
(\ref{eqn:NielsenHughes}), or
\be
\beta_0 = - \frac{(-1)^{2S}}{4\pi^2} \,\left[ \Big\{ (2S)^2 \Big\}_\text{para} +
	\Big\{ - \textstyle\frac{1}{3} \Big\}_\text{dia} \, \right] \,,
\label{eqn:beta0diapara}
\ee
for the cases $S=\frac{1}{2}$ and $S=1$, respectively. However, this formula is valid
more generally for any massless field of spin $S$, carrying nonzero Abelian or
non-Abelian charge, and having a $g$-factor of exactly $2$.

Concerning our discussion of paramagnetic dominance the important point about
(\ref{eqn:beta0diapara}) is the following: even though for all spins $S \geq \frac{1}{2}$
the paramagnetic contribution to $\beta_0$ is larger than the diamagnetic one which comes
with the opposite sign, the total sign of $\beta_0$ nevertheless alternates with $S$.
This results from the overall factor $(-1)^{2S}$ which can be seen as a consequence of
the spin-statistic theorem or of the Feynman rule requiring an extra minus sign for every
fermion loop. Without this extra factor, QED would be asymptotically free since both in
Yang-Mills theory \emph{and in QED} the ``para'' contribution to $\beta_0$ overrides the
``dia'' one.

Asymptotic freedom can be understood in an elementary way by viewing the vacuum state of
the quantum field theory under consideration as a magnetic medium and analyzing the
response of this medium to an external magnetic field
\cite{Nielsen,Johnson,huang-book,Gottfried-Weisskopf}. Besides the (electromagnetic or
color) fields $\mathbf{E}$, $\mathbf{B}$ it is then useful to introduce also
$\mathbf{D} = \mathbf{E} + \mathbf{P}$ and $\mathbf{H} = \mathbf{B} - \mathbf{M}$ with
the polarization $\mathbf{P}$ and the magnetization $\mathbf{M}$, respectively \cite{HE}.
In terms of the corresponding effective Lagrangian
$\mathcal{L}_\text{eff}(\mathbf{E},\mathbf{B})$,
\be
D_i = \frac{\p \mathcal{L}_\text{eff}}{\p E_i} \;, \qquad
H_i = - \frac{\p \mathcal{L}_\text{eff}}{\p B_i} \;.
\ee
Introducing electric and magnetic susceptibilities $\chi_\text{el}$ and $\chi_\text{mag}$
by $\mathbf{P} = \chi_\text{el}\, \mathbf{E}$ and $\mathbf{M} = \chi_\text{mag}\,
\mathbf{H}$, and the permeabilities $\varepsilon = 1 + \chi_\text{el}$ and
$\mu = 1 + \chi_\text{mag}$, we have then $\mathbf{D} = \varepsilon(\mathbf{E},\mathbf{B})
\, \mathbf{E}$ and $\mathbf{H} = \mu(\mathbf{E},\mathbf{B})^{-1}\, \mathbf{B}$. The
field dependence of $\varepsilon$ and $\mu$ (which are tensors in general) is determined
by $\mathcal{L}_\text{eff}$. As usual a medium is called diamagnetic if $\chi_\text{mag}
<0$, and paramagnetic if $\chi_\text{mag}>0$.

Up to here the setting is exactly the same as in condensed matter physics. A special
feature of Lorentz invariant quantum field theories on Minkowski space is that
$\varepsilon \mu = 1$, or $\varepsilon(\mathbf{E},\mathbf{B}) =
\mu(\mathbf{E},\mathbf{B})^{-1}$. Hence, when $\varepsilon$ and $\mu$ are close to unity
we have approximately $\chi_\text{el} \approx - \chi_\text{mag}$.

A homogeneous, isotropic medium, the vacuum of some quantum field theory, say, is
\emph{screening} electric charges if $\varepsilon > 1$, $\chi_\text{el} > 0$. In the
relativistic case this implies $\mu < 1$, $\chi_\text{mag} < 0$, and so the vacuum is a
\emph{diamagnetic medium}.

If, instead, the medium is \emph{antiscreening} electric charges we have
$\varepsilon < 1$, $\chi_\text{el} < 0$, and Lorentz invariance implies $\mu > 1$,
$\chi_\text{mag} > 0$. The vacuum state represents a \emph{paramagnetic medium} then.

Here we used the usual terminology of calling a medium dia- (para-) magnetic when
$\mu < 1$ ($\mu > 1$). We stress that a priori this notion has little or nothing to do
with our earlier discussion of dia- vs.\ paramagnetic \emph{interactions}. It is only in
the nonrelativistic theory of standard magnetic materials that ``paramagnetic dominance''
leads to a ``paramagnetic medium''. In the present generalized context, however,
\emph{the vacuum state of a quantum field theory can behave as a diamagnetic medium even
though paramagnetic interactions dominate}.

This possibility is closely related to the point we made about the factor $(-1)^{2S}$ in
the formula (\ref{eqn:beta0diapara}) for $\beta_0$. In fact, if one computes
$\mathcal{L}_\text{eff}$ for the field theory this formula applies to, and determines the
field dependent magnetic susceptibility from it, the result reads
\cite{Nielsen,Johnson,huang-book}:
\be
\chi_\text{mag}(B) = - \frac{1}{2}\,\beta_0\, g^2\, \ln
	\left( \frac{\Lambda^2}{gB} \right) \,.
\label{eqn:chiYM}
\ee
Here $\Lambda$ is a UV cutoff, and we employ a normalization such that
$\chi_\text{mag}(B=\Lambda^2/g) = 0$. Lowering $B$ below $\Lambda^2/g$ we integrate out
the modes with eigenvalues in the interval $[gB,\Lambda^2]$. This renders
$\chi_\text{mag}$ nonzero whereby its sign is correlated with the sign of $\beta_0$:
\be
\begin{split}
\beta_0 > 0 \quad &\Rightarrow\quad \chi_\text{mag} < 0 \,,\; \text{diamagnetic medium}\\
\beta_0 < 0 \quad &\Rightarrow\quad \chi_\text{mag} > 0 \,,\; \text{paramagnetic medium}
\end{split}
\ee

As a consequence, since $\beta_0^\text{QED} > 0$ the vacuum of QED is a diamagnetic
medium, while by virtue of $\beta_0^\text{YM} < 0$ the vacuum state of non-Abelian gauge
bosons is a paramagnetic one. But in both cases we have ``paramagnetic dominance'' as far
as the relative strength of the two interactions is concerned!

Again, the perhaps unexpected diamagnetism of the QED vacuum is due to the extra minus
sign for fermions. Note that in scalar electrodynamics we have $\beta_0 > 0$, too. For
$S=0$ we loose the minus sign from the statistics now, but since the scalar has no
paramagnetic interaction at all and shows ``diamagnetic dominance'', the bracket in
(\ref{eqn:beta0diapara}) also changes its sign (relative to the spinor case). Thus, in
scalar electrodynamics, we have the perhaps less surprising association of diamagnetic
material properties to diamagnetic interactions.

\section{Paramagnetic dominance and Asymptotic Safety in QEG}
\label{sec:QEG}
In this section we investigate Asymptotic Safety in Quantum Einstein Gravity.
Using the gravitational average action \cite{mr} many different truncations and
models have been analyzed \cite{NJP,livrev,reviews}. These studies all
share one central result: the existence of a non-Gaussian fixed point (NGFP), the crucial
prerequisite for a theory to be asymptotically safe. However, the most obvious question
remained open: what is the physical reason for this fixed point to appear? Why do all
those many independent computations conspire to give the same result? And, could the NGFP
get ``destroyed'' in some way? In order to approach this problem we shall avail ourselves
of the same concept as in the previous section, based on the generalized dia- and
paramagnetic interactions of quantum fluctuations with their background. The analogy
between gravity and Yang-Mills theory is made clear in the following.

In the framework of the gravitational average action the dynamical metric $\gamma_\mn$ is
written as a sum of a fixed background metric $\bar{g}_\mn$ and the fluctuations $h_\mn$,
i.e.\ $\gamma_\mn = \bar{g}_\mn + h_\mn$. For its expectation value we have analogously
$g_\mn \equiv \langle\gamma_\mn\rangle = \bar{g}_\mn + \bar{h}_\mn$ with
$\bar{h}_\mn \equiv \langle h_\mn\rangle$. Since $\bar{g}_\mn$ and $\bar{h}_\mn$ enter
the cutoff and gauge fixing terms separately \cite{mr}, the average action
$\Gamma_k[\bar{h}_\mn;\bar{g}_\mn]$, or equivalently
$\Gamma_k[g_\mn,\bar{g}_\mn] \equiv \Gamma_k[g_\mn-\bar{g}_\mn;\bar{g}_\mn]$, depends on
two independent tensor fields.

\piccaption{Schematic diagram of the metric fluctuations coupling to the background.
\label{fig:QEG-Feyn}}
\pichskip{2em}
\parpic[l]{
\fcolorbox{white}{white}{
	\begin{picture}(136,70) (100,-90)
    \SetWidth{0.8}
    \SetColor{Black}
    \Gluon(111,-41)(159,-41){5}{4}
    \SetWidth{1.0}
    \Vertex(159,-40){2.236}
    \Text(164,-92)[lb]{\small{\Black{$\bar{g}_{\mu\nu}$}}}
    \Line[dash,dashsize=2](159,-89)(159,-43)
    \Text(113,-60)[lb]{\small{\Black{$h_{\mu\nu}$}}}
    \Text(195,-60)[lb]{\small{\Black{$h_{\mu\nu}$}}}
    \SetWidth{0.8}
    \Gluon(160,-41)(208,-41){5}{4}
  \end{picture}}
}
\noindent
After expanding $\Gamma_k$ in terms of $\bar{h}_\mn$ one encounters interaction terms
between the metric fluctuations and the background field of any order in $\bar{h}_\mn$,
schematically indicated in figure \ref{fig:QEG-Feyn}. In a single-metric truncation where,
by definition, $\Gamma_k[g,\bar{g}]$ has only a trivial $\bar{g}_\mn$-dependence via the
gauge fixing term, a perturbative evaluation of the supertrace in the corresponding FRGE
involves only diagrams with external $\bar{g}_\mn$-lines and $h_\mn$'s propagating inside
loops. Within this class of truncations, it is exclusively the fluctuation-background
interaction that drives all RG effects; self-interactions of the $h_\mn$'s play no
r\^{o}le yet.

From the point of view of the metric fluctuations (``gravitons'') the background geometry
can be regarded as a kind of external ``magnetic'' field, polarizing the quantum vacuum
of the ``$h_\mn$-particles'', and giving rise to an induced field energy and a
corresponding susceptibility. Therefore, a separation of dia- and paramagnetic mechanisms
by disentangling kinetic from ultralocal alignment effects is natural also here.

\subsection{Einstein-Hilbert truncation: dia- vs.\ paramagnetism}
\label{sec:EHDiaPara}
Next we study QEG within the Einstein-Hilbert truncation. We derive the $\beta$-functions
along the same lines as in \cite{mr}, but having computed the inverse propagator
$\Gamma_k^{(2)}$ we perform the split into the magnetic components and treat them
separately during the entire calculation that follows.

Our specific truncation ansatz consists of the Einstein-Hilbert action with a scale
dependent Newton and cosmological coupling constant, $G_k$ and $\Lambda_k$,
respectively, plus a gauge fixing and a ghost action:
$\Gamma_k = \Gamma_k^\text{EH} + \Gamma_k^\text{gf} + \Gamma_k^\text{gh}
\equiv \breve{\Gamma}_k + \Gamma_k^\text{gh}$, where $\breve{\Gamma}_k$ is given by
\be
\begin{split}
\breve{\Gamma}_k[g,\bar{g}] =\, &\frac{1}{16\pi G_k} \int\ddx\, \sg \left( -R[g] 
		+ 2\Lambda_k \right)\\
	& + \frac{1}{32\pi G_k} \int\ddx\, \sgb\,\bar{g}^\mn \big(\mathcal{F}_\mu^{\alpha\beta}
		g_{\alpha\beta} \big) \big(\mathcal{F}_\nu^{\rho\sigma}	g_{\rho\sigma} \big) \,.
\label{eqn:Einstein-Hilbert}
\end{split}
\ee
Here we use the harmonic coordinate condition, i.e.\ 
$\mathcal{F}_\mu^{\alpha\beta} \equiv \delta_\mu^\beta\,\gb^{\alpha\gamma}
\bar{D}_\gamma - \frac{1}{2} \gb^{\alpha\beta}\bar{D}_\mu$.
With this gauge choice the Faddeev-Popov operator $\mathcal{M}$ reads
$\mathcal{M}[g,\bar{g}]^\mu_{~\nu} = \gb^{\mu\rho} \gb^{\sigma\lambda} \bar{D}_\lambda
	(g_{\rho\nu}D_\sigma + g_{\sigma\nu}D_\rho) - \gb^{\rho\sigma} \gb^{\mu\lambda}
	\bar{D}_\lambda g_{\sigma\nu}D_\rho \,$.
Since we do not want to determine the running of the ghost sector here,
$\Gamma_k^\text{gh}$ coincides with the classical ghost action. This leads to a
decomposed FRGE, with one trace for the metric fluctuations and another one for the
ghosts:
\be
\begin{split}
\p_t\Gamma_k[g,\bar{g}] = \frac{1}{2}\, &\Tr\left[\left(\breve{\Gamma}_k^{(2)}[g,\bar{g}]
	+ \Rk^\text{grav}[\bar{g}] \right)^{-1} \p_t\Rk^\text{grav}[\bar{g}] \right]\\
	- &\Tr \left[ \left( -\mathcal{M}[g,\bar{g}] + \Rk^\text{gh}[\bar{g}]\right)^{-1}
	\p_t\Rk^\text{gh}[\bar{g}]\right] \,.
\label{eqn:FRGE-QEG}
\end{split}
\ee
Here $\Rk^\text{grav}[\bar{g}]\propto k^2 R^{(0)}(-\bar{D}^2/k^2)$, and similarly for
$\Rk^\text{gh}[\bar{g}]$.\footnote{See \cite{oliver,frank1} and in particular the
summary in \cite{CPR} for a discussion of other types of cutoff operators
$\Rk^\text{grav}$. It is by now well established that for pure gravity in the
Einstein-Hilbert truncation the qualitative features of the RG flow are insensitive
to the precise definition of $\Rk$. Replacing, for instance, $\bar{D}^2$ in the argument
of $R^{(0)}$ by a nonminimal operator $\bar{D}^2+\text{curvature}$ has no significant
effect.} For the projection it is sufficient to set $\gb = g$ after having determined the second
functional derivative $\breve{\Gamma}_k^{(2)}$. This leads to the nonminimal operator
\be
\left(\breve{\Gamma}_k^{(2)}[g,\gb]\right)^\mn_{~~\rs}\Big|_{g=\gb} = \frac{1}{32\pi G_k}
	\,\Big( -\bar{K}^\mn_{~~\rs} \bar{D}^2 + \bar{U}^\mn_{~~\rs} \Big) \,,
\label{eqn:Gamma2Grav}
\ee
with $\bar{K}^\mn_{~~\rs}$ and $\bar{U}^\mn_{~~\rs}$ as introduced in eqs.\
(\ref{eqn:GravOperator}), (\ref{eqn:UGrav}). Furthermore, the Faddeev-Popov operator
assumes a similar nonminimal form, involving the covariant Laplacian $-\bar{D}^2$ and a
potential term:
\be
-\mathcal{M}[g,\bar{g}]^\mu_{~\nu}\Big|_{g=\gb}= \delta^\mu_\nu\Big(-\bar{D}^2 
	-\frac{1}{d}\,\bar{R}\Big) \,.
\label{eqn:Faddeev-Popov}
\ee

At this point we perform the separation into the different ``magnetic'' components.
Contributions coming from the first term in (\ref{eqn:Gamma2Grav}) are referred to as
\emph{diamagnetic}, those from the second term as \emph{paramagnetic}. Similarly, the
first part of the Faddeev-Popov operator (\ref{eqn:Faddeev-Popov}) gives rise to
\emph{ghost-diamagnetic} interactions, while the second one induces
\emph{ghost-paramagnetic} effects.

Next we decompose $\bar{h}_\mn$ into a trace plus a traceless part, and we
assume that $\gb_\mn$ corresponds to a $d$-sphere, which simplifies the computation but
is general enough to identify the terms of our truncation. Then the curvature scalar $R$
is no longer a function but rather a numerical constant depending on the radius of the
$d$-sphere. With these assumptions the FRGE reads
\be
\begin{split}
\p_t\Gamma_k[g] =\, &\Tr{\!}_\text{T}\left[\mathcal{N}(\mathcal{A} + C_\text{T}R)^{-1}
	\right]
	+ \Tr{\!}_\text{S}\left[ \mathcal{N}(\mathcal{A} + C_\text{S} R)^{-1} \right] \\
	&- 2 \Tr{\!}_\text{V}\left[ \mathcal{N}_0(\mathcal{A}_0 + C_\text{V} R)^{-1} \right] \,,
\end{split}
\label{eqn:FRGE-QEG2}
\ee
where the traces $\displaystyle{\Tr{\!}_\text{T}}$, $\displaystyle{\Tr{\!}_\text{S}}$ and
$\displaystyle{\Tr{\!}_\text{V}}$ refer to symmetric traceless tensors, scalars and
vectors, respectively. The constants $C_\text{T}$, $C_\text{S}$ and $C_\text{V}$ are
given by
\be
C_\text{T} \equiv \frac{d(d-3)+4}{d(d-1)} \;, \quad C_\text{S} \equiv \frac{d-4}{d} \quad
	\text{and}\quad C_\text{V} \equiv -\frac{1}{d} \;.
\ee
In (\ref{eqn:FRGE-QEG2}) we also introduced the following functions of the Laplacian
$D^2$,
\begin{align}
\mathcal{A} &\equiv -D^2 + k^2 \RN(-D^2/k^2) - 2\Lambda_k \,, \\
\mathcal{N} &\equiv ( 1 - \textstyle{\frac{1}{2}} \eta_N ) k^2 \RN(-D^2/k^2) 
	+ D^2{\RN}'(-D^2/k^2) \,,
\end{align}
and their ghost counterparts $\mathcal{A}_0 \equiv \mathcal{A}\big|_{\Lambda_k=0}$
and $\mathcal{N}_0 \equiv \mathcal{N}\big|_{\eta_N=0}\,$, with
the anomalous dimension of Newton's constant $\eta_N \equiv \p_t\ln G_k$.

The crucial point is that on the RHS of (\ref{eqn:FRGE-QEG2}) the denominators in each of
the three traces are of the same form, and that the paramagnetic contribution is given
entirely by a term $\propto C R$. Therefore, when we perform the expansion
\be
(\mathcal{A}+CR)^{-1} = \mathcal{A}^{-1} - C\mathcal{A}^{-2}\,R+\ord(R^2) \,,
\ee
we identify the term $\mathcal{A}^{-1}$ as diamagnetic and the one proportional to $R$ as
paramagnetic.

The next steps are exactly the same as in \cite{mr}. This finally yields two differential
equations describing the scale dependence of $G_k$ and $\Lambda_k$, or of the
corresponding dimensionless couplings, $g_k \equiv k^{d-2}G_k$ and
$\lambda_k \equiv k^{-2} \Lambda_k$, respectively.

For the cosmological constant we find the flow equation
\be
\begin{split}
\p_t \lambda_k = \beta_\lambda(g_k,\lambda_k) \equiv\, &\big[\eta_N(g_k,\lambda_k)-2\big]
	\lambda_k + 2\pi g_k (4\pi)^{-d/2} \Big[ 2d(d + 1)\, \Phi_{d/2}^1(-2\lambda_k) \\
	&-d(d + 1)\, \eta_N(g_k,\lambda_k)\, \widetilde\Phi_{d/2}^1(-2\lambda_k)
	- 8d\, \Phi_{d/2}^1(0)\Big] \,,
\end{split}
\label{eqn:betalambda}
\ee
where the threshold functions $\Phi$ and $\widetilde\Phi$ are given by eqs.\
(\ref{eqn:Phi}) and (\ref{eqn:PhiTilde}) of the previous section. The separation rule
outlined above is now used to identify the different magnetic contributions to the
anomalous dimension. We obtain the result in the familiar form
\be
\eta_N(g,\lambda) = \frac{g \, B_1(\lambda)}{1-g \, B_2(\lambda)} \,.
\label{eqn:etaN}
\ee
This representation of $\eta_N$ contains two functions of the cosmological constant, both
with ``dia'' and ``para'' contributions, from both the gravitons and the Faddeev-Popov
ghosts. In the numerator of (\ref{eqn:etaN}) we have
\be
\begin{split}
B_1(\lambda) \equiv\, \frac{1}{3}\, &(4\pi)^{1-\frac{d}{2}} \Bigg[ \bigg\{
	d(d+1)\, \Phi_{d/2-1}^1(-2\lambda) \bigg\}_\text{dia}
	+ \bigg\{ -4d\, \Phi_{d/2-1}^1(0) \bigg\}_\text{ghost-dia} \\
	&\qquad + \bigg\{ -6d(d-1)\, \Phi_{d/2}^2(-2\lambda) \bigg\}_\text{para}
	+ \bigg\{-24\, \Phi_{d/2}^2(0) \bigg\}_\text{ghost-para} \,\Bigg] \,,
\end{split}
\label{eqn:QEGB1}
\ee
and similarly in the denominator:
\be
\begin{split}
B_2(\lambda) \equiv -\frac{1}{6}\, (4\pi)^{1-\frac{d}{2}}\Bigg[
	& \bigg\{ d(d+1) \widetilde\Phi_{d/2-1}^1(-2\lambda) \bigg\}_\text{dia} 
	+ \bigg\{ -6d(d-1)\widetilde\Phi_{d/2}^2(-2\lambda)\bigg\}_\text{para}\, \Bigg].
\end{split}
\label{eqn:QEGB2}
\ee
In terms of the anomalous dimension, the RG equation of the dimensionless Newton constant
reads
\be
\p_t g_k = \beta_g(g_k,\lambda_k) \equiv \big[ d-2+\eta_N(g_k,\lambda_k) \big]\, g_k \,.
\label{eqn:betag}
\ee

As eq.\ (\ref{eqn:etaN}) suggests, the specific form of $\eta_N$ displays two parts of
different relevance: the numerator $g\, B_1(\lambda)$ determines the qualitative behavior
of $\eta_N$, in particular it decides on the overall sign of $\eta_N$. By contrast, the
denominator $1-g\, B_2(\lambda)$ plays the r\^{o}le of a correction term only. It
stems from the differentiated $G_k$ factors inside $\Rk$ on the RHS of the FRGE. Due to
the singularity it causes at $g=1/B_2(\lambda)$ it delimits the $g$-$\lambda$ theory
space \cite{frank1}, but away from this boundary singularity, where the calculation can
be trusted, it does not lead to qualitative changes of the leading order behavior given
by the numerator.

Since our analysis focuses on the sign of $\eta_N$, the important
features are contained in $g\, B_1(\lambda)$ alone. In order to investigate the influence
of the various magnetic effects it is thus sufficient to expand $\eta_N$ in powers of $g$,
$\eta_N = g\, B_1(\lambda) + \ord(g^2)$, and to retain the term linear in $g$
only:\footnote{Despite this expansion in $g$, we are still aiming at a nonperturbative
renormalization of QEG, meaning that the continuum limit is taken at an NGFP with
$g_*\neq 0$ rather than the trivial fixed point of perturbation theory, $g_*=0$. If the
NGFP, in some scheme, has a small but nonzero $g_* B_2(\lambda_*)$ it might well be
possible to find it in a small coupling expansion. However, the latter may not be
confused with ``perturbation theory'' in the sense of ``perturbative renormalization''.}
\be
\begin{split}
\eta_N(g,\lambda &) =\, \frac{1}{3}\, (4\pi)^{1-\frac{d}{2}}\Bigg[ \bigg\{
	d(d+1)\, \Phi_{d/2-1}^1(-2\lambda) \bigg\}_\text{dia}
	+ \bigg\{ -4d\, \Phi_{d/2-1}^1(0) \bigg\}_\text{ghost-dia} \\
	& + \bigg\{ -6d(d-1)\, \Phi_{d/2}^2(-2\lambda) \bigg\}_\text{para}
	+ \bigg\{-24\, \Phi_{d/2}^2(0) \bigg\}_\text{ghost-para} \,\Bigg] g
	+ \ord\left( g^2 \right).
\end{split}
\label{eqn:etaNexpansion}
\ee
Already at the level of (\ref{eqn:etaNexpansion}) we can make an important observation if
we take into account that the $\Phi$'s are strictly positive functions: \emph{The
graviton's paramagnetic part as well as both ghost contributions (dia- and paramagnetic)
drive $\eta_N$ towards negative values, while the graviton's diamagnetic term has the
opposite sign and tries to make $\eta_N$ positive}.

The numerical prefactor $6d(d-1)$ of the (graviton-) paramagnetic term, however, is
larger than the one of the diamagnetic part, $d(d+1)$, for any $d > 1.4$, suggesting that
the overall sign of $\eta_N$ is governed by the three non-graviton-diamagnetic effects.
Of course, the validity of this hypothesis has to be checked for a generic cutoff shape
function $\RN$, which enters (\ref{eqn:etaNexpansion}) through the threshold functions,
see eqs.\ (\ref{eqn:Phi}) and (\ref{eqn:PhiTilde}). Later on we shall indeed demonstrate
the universality of our findings, and in particular that the conjecture about $\eta_N$
being negative due to ``paramagnetic dominance'' is actually true, by employing a whole
class of cutoff functions $\RN$.

Why is the sign of $\eta_N$ so crucial? By virtue of the definition $\eta_N \equiv k\p_k
\ln G_k$, gravitational antiscreening, i.e.\ increasing $G_k$ for decreasing scale $k$,
amounts to $\eta_N<0$. Furthermore, our main interest consists in finding a non-Gaussian
fixed point $(g_*,\lambda_*)$, as it is the fundamental ingredient for the Asymptotic
Safety scenario. In the nontrivial case ($g_*\neq 0$) a fixed point requires, by eq.\
(\ref{eqn:betag}), that $\eta_N(g_*,\lambda_*) = 2-d$. Thus, for any $d>2$, \emph{an NGFP
can occur only if $\eta_N$ is negative}.

With regard to (\ref{eqn:etaNexpansion}) we conclude that an asymptotically safe world
needs the gra\-vi\-ton-di\-a\-mag\-ne\-tic effect to lose against the three other ones.
In the following we show that this is actually the case.

\subsection[Paramagnetic dominance: \texorpdfstring{$\eta_N$}{nN} to leading order 
	in \texorpdfstring{$g$}{g}]{Paramagnetic dominance: $\bm{\eta_N}$ to leading order
	in $\bm{g}$}
\label{sec:etaNtoLO}
In this subsection we investigate under which conditions the NGFP forms. We begin with a
general discussion for arbitrary dimension $d$, before we turn to the special cases of
$4$, $3$, and $2+\epsilon$ dimensions.

In order to compare the relative size of the different magnetic contributions to
$\eta_N$ we need to resort to a particular cutoff. A simple choice is the ``optimized''
shape function \cite{opt}, $\RN(z)=(1-z)\Theta(1-z)$, which allows for analytical results.
Additionally, the main findings are checked afterwards using a family of exponential
shape functions. This is necessary to show that the conclusions obtained with the
optimized $\RN$ are universal.

For the optimized cutoff $\eta_N$ assumes the explicit form
\be
\begin{split}
\eta_N(g,\lambda) =\, \frac{1}{3}\,(4\pi)^{1-\frac{d}{2}} &\, \frac{1}{\Gamma(d/2)}\,
	\Bigg[ \bigg\{ \frac{d(d+1)}{1-2\lambda} \bigg\}_\text{dia}
	+ \bigg\{ -4d \bigg\}_\text{ghost-dia} \\
	&+ \bigg\{ -\frac{12(d-1)}{(1-2\lambda)^2} \bigg\}_\text{para}
	+ \bigg\{ -\frac{48}{d} \bigg\}_\text{ghost-para} \,
	\Bigg] g + \ord(g^2) \,.
\end{split}
\label{eqn:etaNopt}
\ee
To figure out the relative importance of the four terms in (\ref{eqn:etaNopt}) we first
set $\lambda=0$, and consider $\lambda \neq 0$ subsequently.

\begin{figure}
\begin{minipage}{.49\columnwidth}
	\flushleft
	\includegraphics[width=\columnwidth]{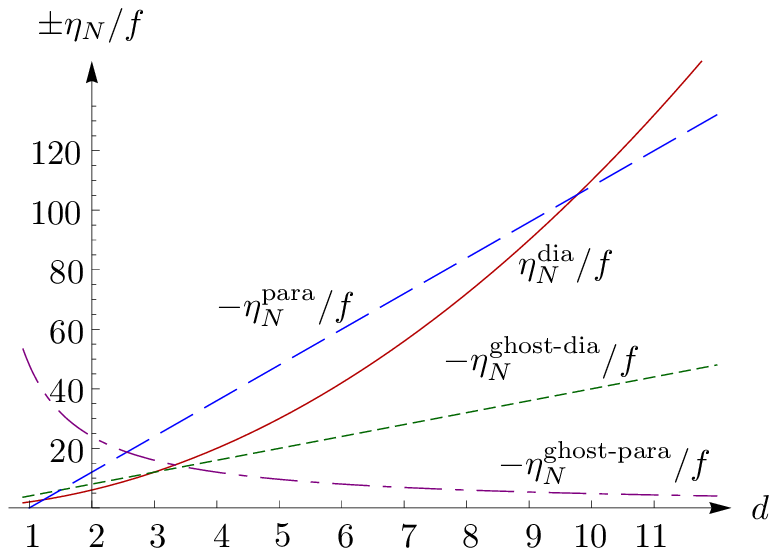}
\end{minipage}
\hfill
\begin{minipage}{.49\columnwidth}
	\flushright
	\includegraphics[width=\columnwidth]{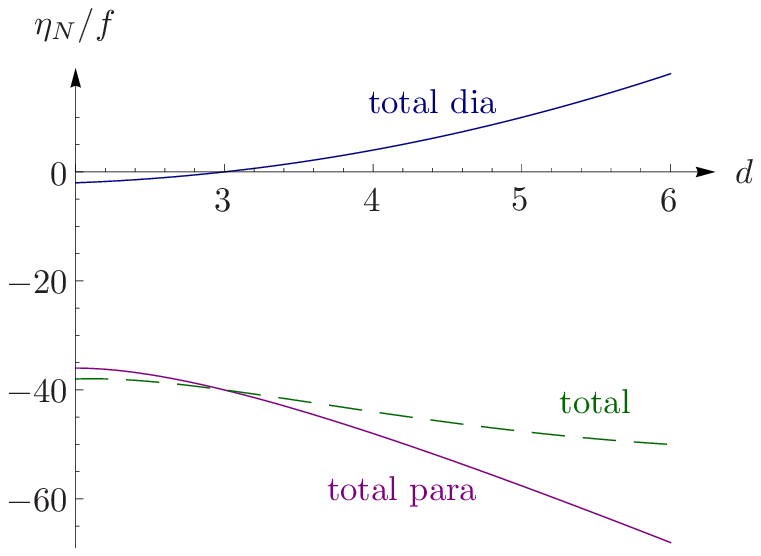}
\end{minipage}
\caption{Relative size of the various magnetic contributions to $\eta_N/f$, where $f$
		denotes the common factor $\frac{1}{3\Gamma(d/2)}(4\pi)^{1-\frac{d}{2}}\,g \,$.
		On the left panel the absolute values of all four contributions (dia, para, ghost-dia
		and ghost-para) are shown. (Note that $\eta_N^\text{dia}$ has an opposite sign.)
		The right panel combines graviton-dia and ghost-dia as well as graviton-para and
		ghost-para parts. Here the total paramagnetic term overbalances the diamagnetic one,
		rendering the sum $\eta_N=\eta_N^\text{total dia}+\eta_N^\text{total para}$ negative
		(dashed line).}
	\label{fig:etaNtot}
\end{figure}

Comparing the absolute values of the four curly brackets in (\ref{eqn:etaNopt}), for
$\lambda=0$, the most important result is that for any $d \lesssim 9.8$ there is indeed
always a contribution present which is larger than the diamagnetic one. For $2.6 \lesssim
d \lesssim 9.8$ it is the graviton-paramagnetic part that provides the largest
contribution, while for $d \lesssim 2.6$ the ghost-paramagnetic effect is most important,
see left panel of figure \ref{fig:etaNtot}. Only for $d \gtrsim 14.4$ the
graviton-diamagnetic term would be large enough to win against the sum of the three other
ones and flip the sign of $\eta_N$. In $d=4$ for instance, we find the hierarchy
\be
\big\{|-36|\big\}_\text{para} > \big\{|+20|\big\}_\text{dia} >
	\big\{|-16|\big\}_\text{ghost-dia} > \big\{|-12|\big\}_\text{ghost-para} \,.
\ee
This confirms that the sign of $\eta_N$ is indeed determined by the three
non-gra\-vi\-ton-dia\-mag\-ne\-tic contributions.

It turns out particularly instructive to combine the graviton-para- and ghost-para-terms
in a total paramagnetic contribution, and similarly in the diamagnetic case. In this way
we obtain from (\ref{eqn:etaNopt}) at $\lambda=0$:
\be
\eta_N(g,0) =\, \frac{1}{3\Gamma(d/2)}\, (4\pi)^{1-\frac{d}{2}} \Big[ \,
	\big\{ d(d-3) \big\}_\text{total dia} + \big\{ -12(d-1)-48/d \big\}_\text{total para}\,
	\Big] g	+ \ord(g^2).
\label{eqn:etaNtot}
\ee
While the total paramagnetic part is always negative, we observe a sign change at $d=3$
in the total diamagnetic component. Thus, for $d<3$, the latter no longer counteracts the
paramagnetic interactions, but rather amplifies their effect of making $\eta_N$ negative.
This sign flip at $d=3$ is independent of the cutoff, it holds for any choice of $\RN$.

The relative total contributions to $\eta_N$ in (\ref{eqn:etaNtot}) are illustrated on
the right panel of figure \ref{fig:etaNtot}.
\begin{figure}
	\begin{minipage}{.47\columnwidth}
		\includegraphics[width=\columnwidth]{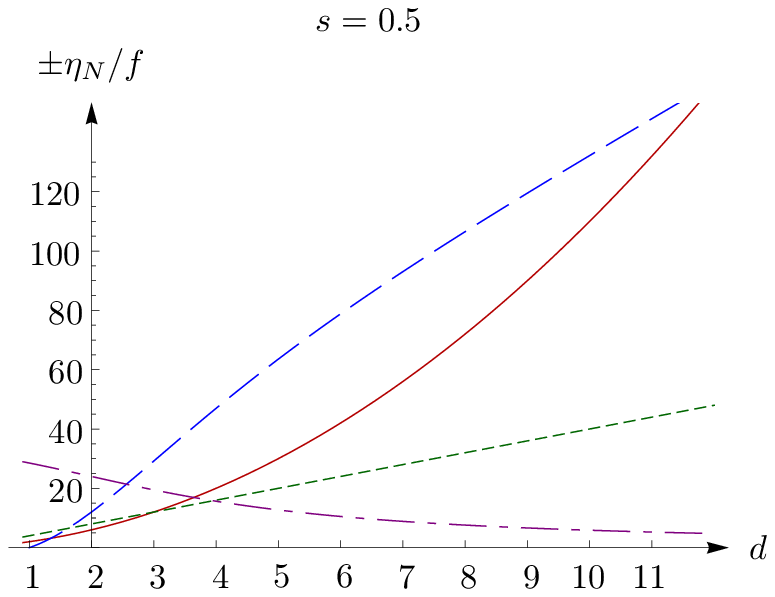}
	\end{minipage}
	\hfill
	\begin{minipage}{.47\columnwidth}
		\includegraphics[width=\columnwidth]{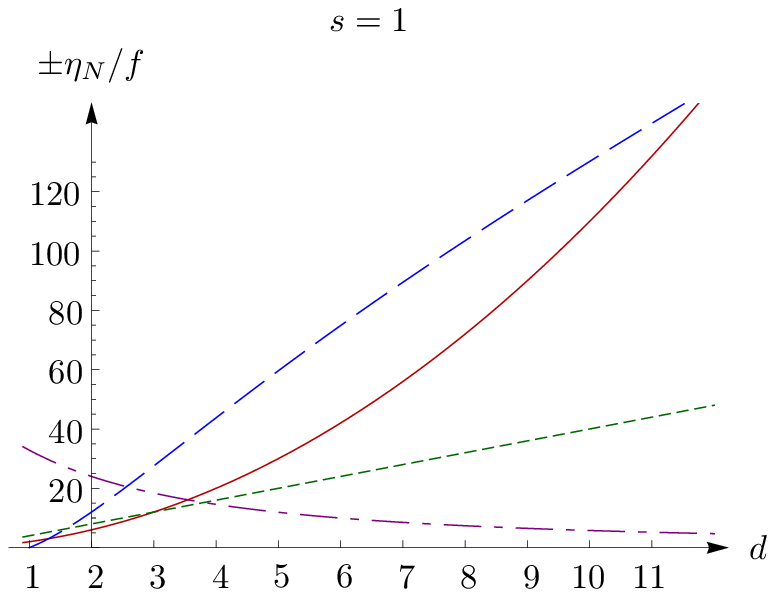}
	\end{minipage}\\[7.5mm]
	\begin{minipage}{.47\columnwidth}
		\includegraphics[width=\columnwidth]{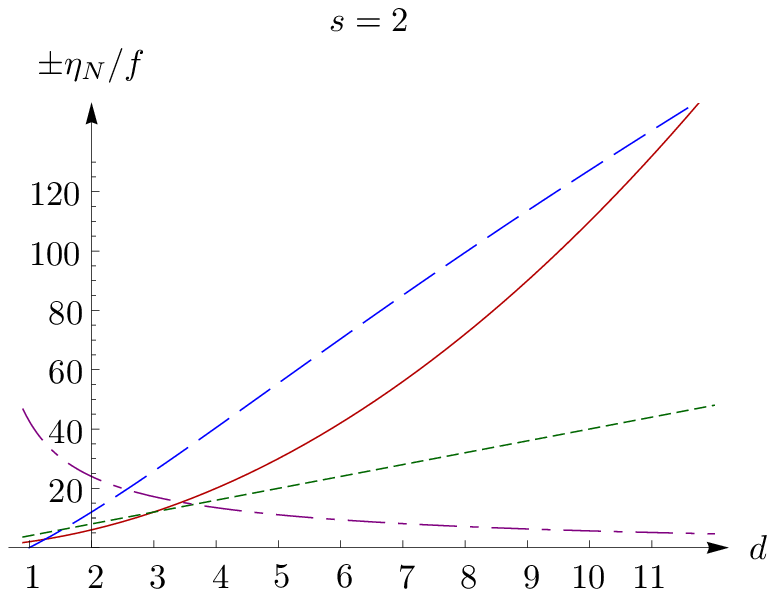}
	\end{minipage}
	\hfill
	\begin{minipage}{.47\columnwidth}
		\includegraphics[width=\columnwidth]{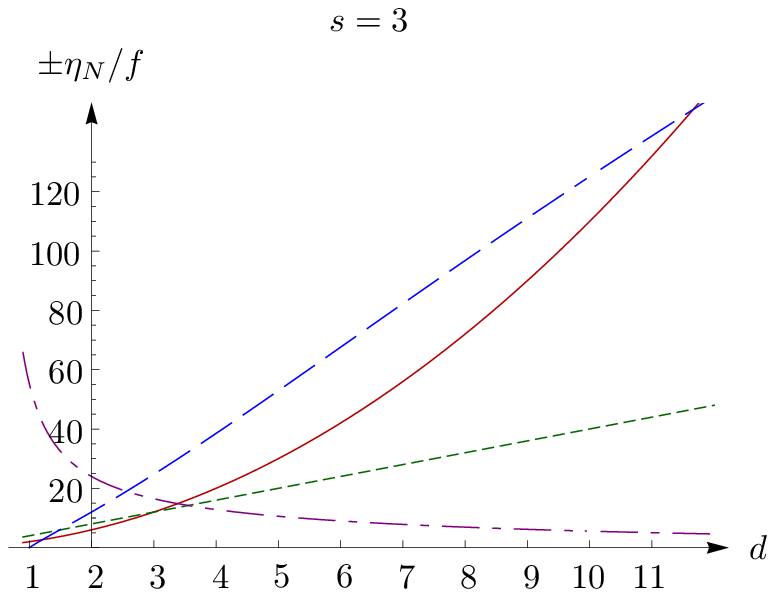}
	\end{minipage}\\[7.5mm]
	\begin{minipage}{.47\columnwidth}
		\includegraphics[width=\columnwidth]{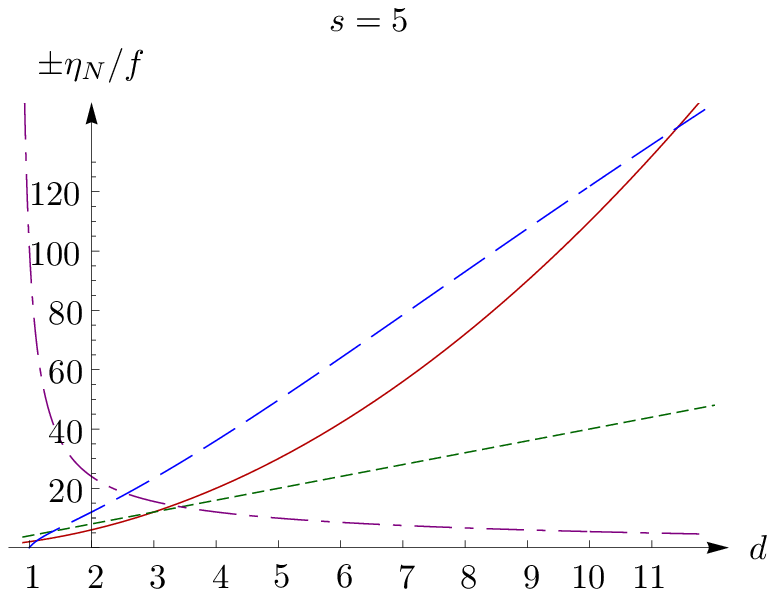}
	\end{minipage}
	\hfill
	\begin{minipage}{.47\columnwidth}
		\includegraphics[width=\columnwidth]{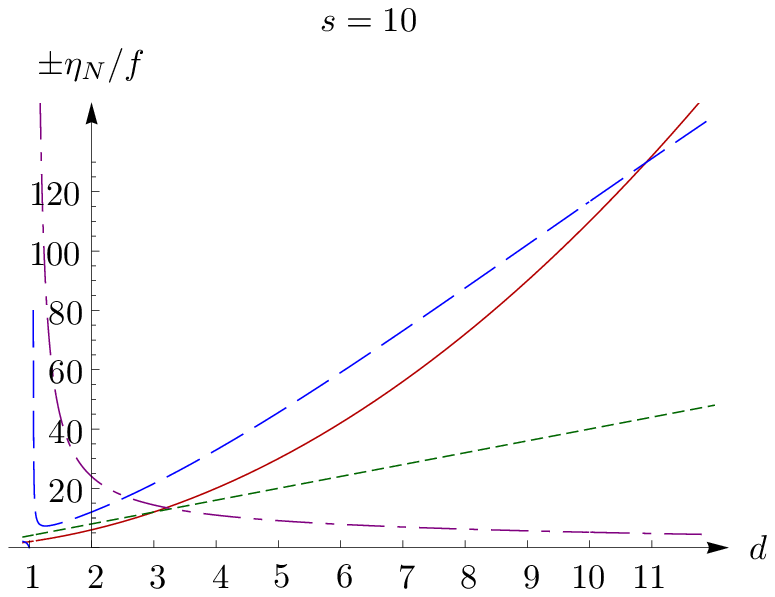}
	\end{minipage}
\caption{Relative size of the various magnetic contributions to the anomalous dimension,
	based on the exponential cutoff $\RN_s$ with $s=0.5,\,1,\,2,\,3,\,5,\,10$. As in figure
	\ref{fig:etaNtot} we normalize each term with their common factor, $f= \frac{1}{3}
	(4\pi)^{1-\frac{d}{2}} \Phi_{d/2-1}^1(0)\, g$. The labeling of the four curves in each
	diagram is analogous to the left panel of figure \ref{fig:etaNtot}.}
\label{fig:etaNExp}
\end{figure}
Here one clearly sees that $\eta_N$ is determined only by the total paramagnetic term, at
least qualitatively. In particular, the negative sign arises only due to paramagnetism.

Next we discuss the general case $\lambda \neq 0$. A careful analysis of the
$\beta$-functions shows that the fixed point value of the cosmological constant,
$\lambda_*$, is positive in most cases. There are only two exceptions: First, for
$d\gtrsim 11.7$ two new fixed points with negative $\lambda_*$ emerge, in addition to the
one with $\lambda_*>0$. We do not discuss such high dimensions here. Second, for
$2<d\lesssim 2.56$ the known fixed point is shifted to negative values for $\lambda_*$.
We will cover this case in detail later on. Therefore, we assume $\lambda>0$ for any
other dimension now.

Let us reconsider eq.\ (\ref{eqn:etaNopt}). The paramagnetic contribution is already
known to be dominant compared to the diamagnetic one for $\lambda=0$.
Going to larger values for $\lambda$ will even enhance this effect due to the factor
$(1-2\lambda)^{-2}$ in the paramagnetic part. Thus, also for general $\lambda$, the
crucial negative sign of $\eta_N$ in the fixed point regime stems from the dominant
paramagnetic terms.

In order to investigate to what extent these findings change when using different cutoffs
we finally repeat our computation of the various contributions to $\eta_N$ for the
one-parameter family of exponential shape functions \cite{frank1,oliver}, $\RN_s(z) =
\frac{sz}{e^{sz}-1}$. The result of this analysis is shown in figure \ref{fig:etaNExp}.
Remarkably, the qualitative picture is almost the same as the one of figure
\ref{fig:etaNtot}. We find paramagnetic dominance for all cutoff functions considered.
The diamagnetic term is too weak to flip the sign of $\eta_N$ if the dimension is not too
large. This demonstrates the universality of our conclusion about the importance of the
paramagnetic interaction terms.

\subsection[Phase portrait and NGFP in \texorpdfstring{$d=4$}{d = 4}]
	{Phase portrait and NGFP in $\bm{d=4}$}
\label{sec:FPQEG4d}
Specializing for 4 dimensions, we shall now investigate the share the dia-/paramagnetic
effects have in the emergence of an NGFP within the Einstein-Hilbert truncation.
First we recall the flow implied by the full $\beta$-functions \cite{mr,frank1},
including all contributions to the anomalous dimension. Then we show that restricting
ourselves to the linear approximation (in $g$) of $\eta_N$ leads to essentially the same
result. Afterwards we perform the same computation, but this time we consider only
paramagnetic terms in $\eta_N$. Finally, we repeat the latter step using diamagnetic
contributions only.

\begin{figure}
\begin{minipage}{0.49\textwidth}
	\centering
	\includegraphics[width=\columnwidth]{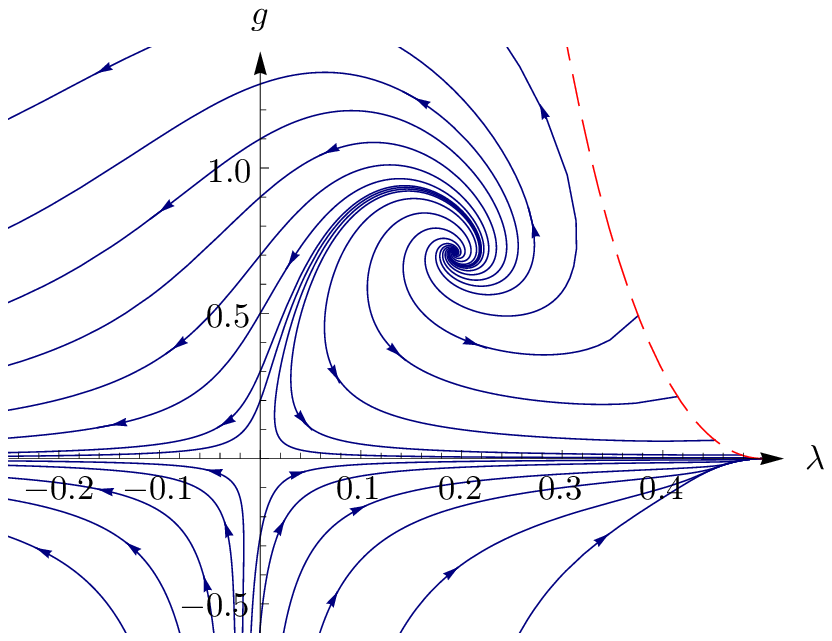}
\end{minipage}
\hfill
\begin{minipage}{0.49\textwidth}
	\centering
	\includegraphics[width=\columnwidth]{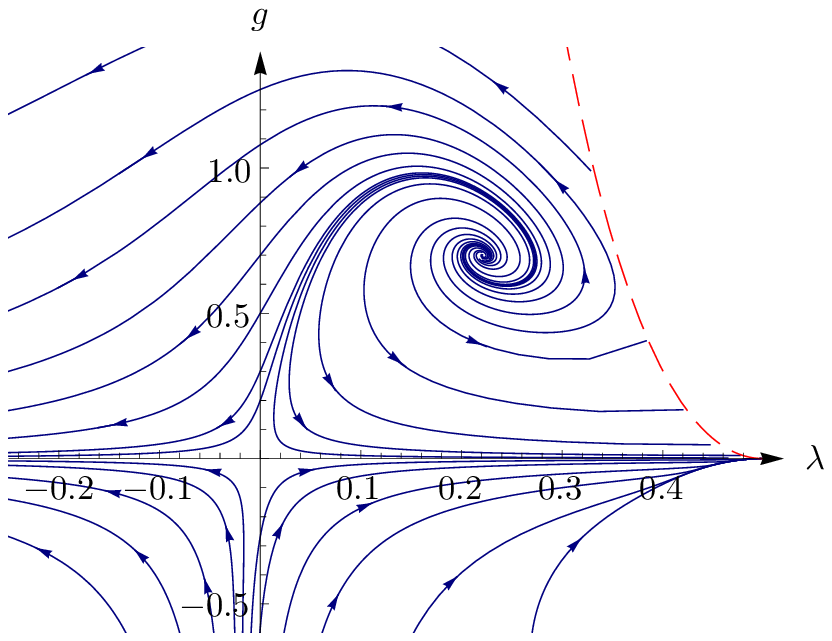}
\end{minipage}	
\caption{Standard phase portrait based on the full $\beta$-functions (left panel) and the
	approximation $\eta_N \approx g\, B_1(\lambda)$, right panel. In each diagram both dia-
	and paramagnetic contributions are retained.}
\label{fig:QEGfull}
\end{figure}

\noindent
\textbf{(i)} We start with the RG equations (\ref{eqn:betalambda}) and (\ref{eqn:betag})
together with the full anomalous dimension (\ref{eqn:etaN}), employing the optimized
cutoff. The resulting phase portrait, obtained by a numerical evaluation, is the well
known one \cite{frank1}; it is depicted on the left panel in figure \ref{fig:QEGfull}.
One finds a Gaussian fixed point in the origin, but also a UV attractive non-Gaussian
fixed point. The dashed curve restricts the domain of the $g$-$\lambda$ theory space since
there the $\beta$-functions diverge. To the left of this boundary all points with positive
Newton's constant are ``pulled'' into the NGFP for $k \rightarrow \infty$.\footnote{The
arrows in the flow diagrams point from the UV to the IR.}
\medskip

\begin{figure}
	\centering
	\includegraphics[width=.75\columnwidth]{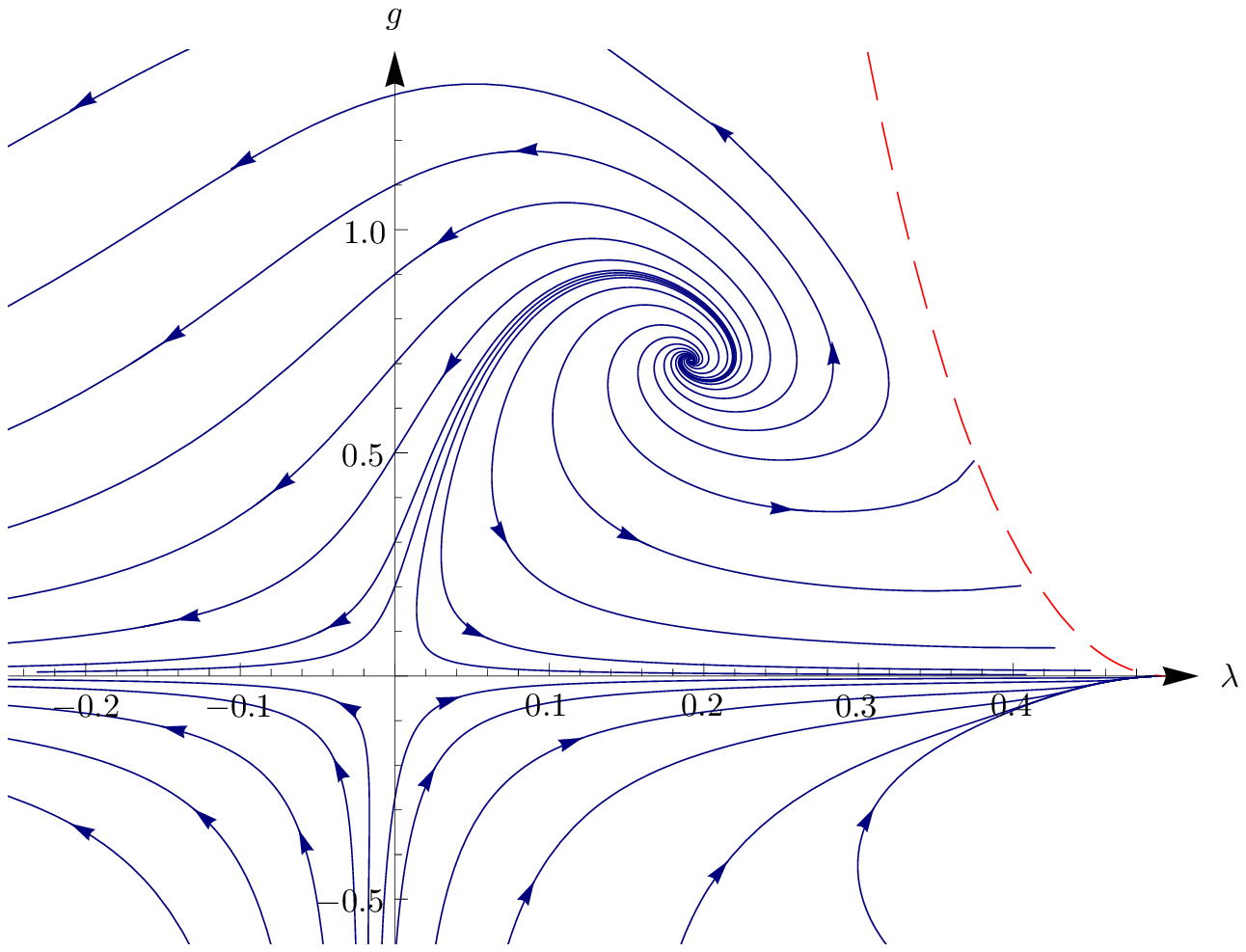}
	\caption{Flow diagram obtained from the \emph{total paramagnetic contributions} to
		$\eta_N$ alone.}
	\label{fig:QEGpara}
\end{figure}

\begin{figure}
	\centering
	\includegraphics[width=.75\columnwidth]{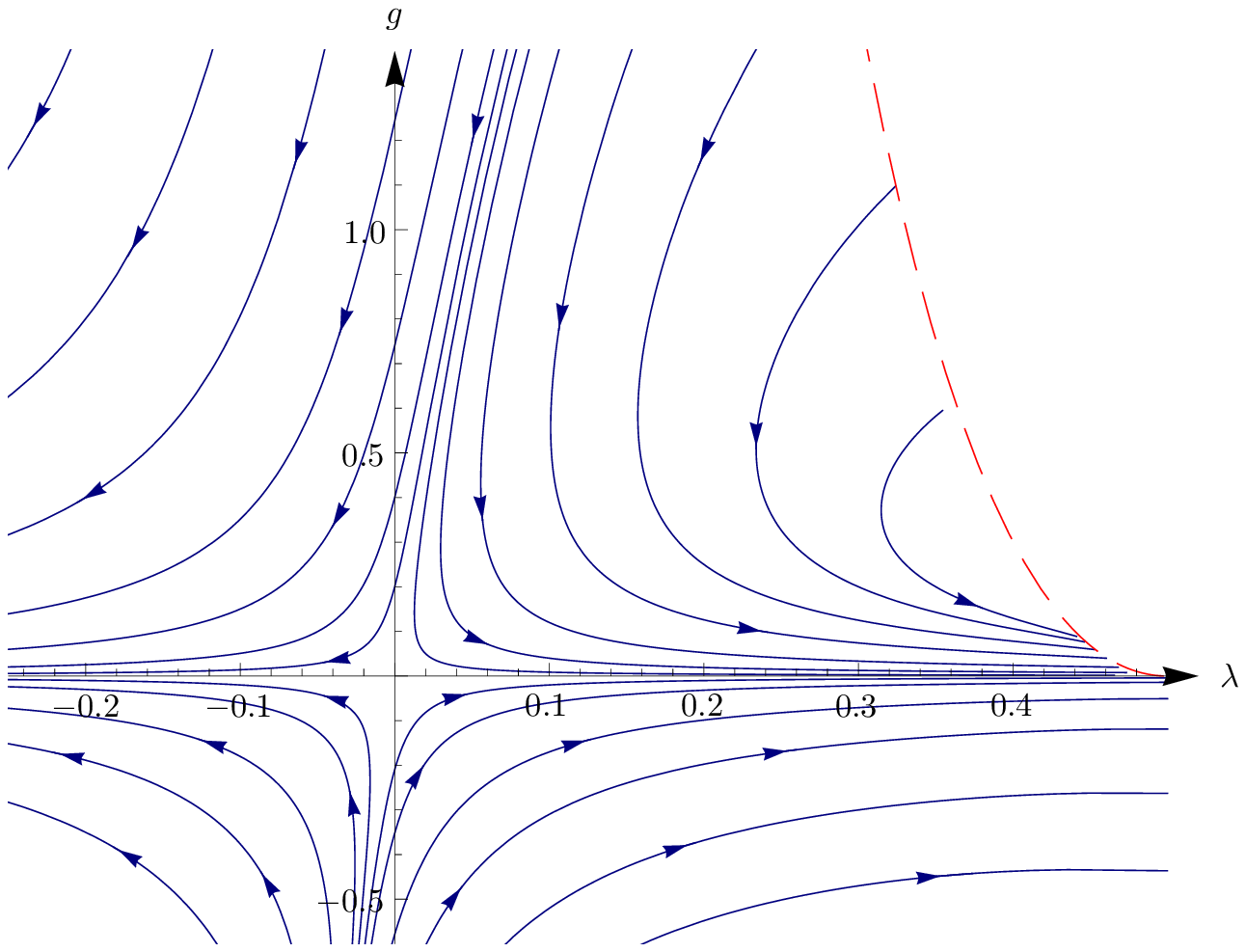}
	\caption{Flow diagram taking into account the \emph{total diamagnetic terms}
		in $\eta_N$ only.}
	\label{fig:QEGdia}
\end{figure}

\noindent
\textbf{(ii)} Now we convince ourselves that the denominator in
$\eta_N = g\, B_1(\lambda)\big/\big(1- g\, B_2(\lambda)\big)$ leads only to qualitatively
inessential modifications of the phase portrait. We solve the RG equations with the
approximate anomalous dimension obtained in leading order of the $g$-expansion,
$\eta_N \approx g\, B_1(\lambda)$, as given in (\ref{eqn:etaNopt}), retaining both dia-
and paramagnetic terms. The resulting phase portrait, shown on the right panel of figure
\ref{fig:QEGfull}, is basically indistinguishable from the exact one on the
left.\footnote{Since we based the computation underlying the right panel of figure
\ref{fig:QEGfull} on the expansion of $\eta_N$ up to first order in $g$,
$\eta_N = g\, B_1(\lambda)$, there is no longer a divergence at $1-g\,B_2(\lambda)=0$.
Nevertheless, we show the dashed line for a better comparison with the left panel. The
same holds for figures \ref{fig:QEGpara} and \ref{fig:QEGdia}.}
Therefore, we may continue with the approximation $\eta_N \approx g\, B_1(\lambda)$.
\medskip

\noindent
\textbf{(iii)} Next we use eq.\ (\ref{eqn:etaNopt}) again, but take into account the
\emph{total paramagnetic contributions only}. Thus $\eta_N$ assumes the simple form
\be
\eta_N^\text{total para}(g,\lambda) = -\frac{1}{\pi} \left[ \frac{3}{(1-2\lambda)^2} + 1
	\right]	g + \ord(g^2) \,,
\label{eqn:etaPara}
\ee
where the first term inside the brackets of (\ref{eqn:etaPara}) is due to the gravitons,
while the ``+1'' stems from the ghosts. We insert this expression into the
$\beta$-functions of $g$ and $\lambda$, and again obtain the flow by a numerical
computation. Figure \ref{fig:QEGpara} displays the resulting phase portrait.

The similarity of this diagram to the phase portrait based on the full $\beta$-functions,
shown in figure \ref{fig:QEGfull}, is truly impressive. We observe that all qualitative
features of the flow are incorporated already in the total paramagnetic terms alone. In
particular, we find the same structure involving a Gaussian fixed point with one
attractive and one repulsive direction, and an NGFP with two UV-attractive
eigendirections. Even the values of the fixed point coordinates and critical exponents
do not change significantly, see Table \ref{tab:FPvalues}.

This is a further demonstration showing that paramagnetism is at the heart of Asymptotic
Safety. The observed behavior is mainly due to the graviton-para term alone: when
omitting the ghost contribution ``+1'' in (\ref{eqn:etaPara}) we find an RG flow similar
to the one of figure \ref{fig:QEGpara}, including the Gaussian and the non-Gaussian fixed
point.
\medskip

\noindent
\textbf{(iv)} At last we perform the same steps as in (iii), but keep only the \emph{total
diamagnetic contributions} to $\eta_N$ in (\ref{eqn:etaNopt}), such that it is given by
\be
\eta_N^\text{total dia}(g,\lambda) =\, \frac{1}{3\pi} \left[ \frac{5}{1-2\lambda} - 4
	\right] g + \ord(g^2) \,,
\ee
where the ``-4'' comes from the ghosts. This anomalous dimension leads to the flow diagram
depicted in figure \ref{fig:QEGdia}. Though the Gaussian fixed point persists, the
structure of the flow is quite different.

It is an important result that the non-Gaussian fixed point has disappeared. This is an
illustration of our statement above that the total diamagnetic term
contributes to $\eta_N$ with the ``wrong'' sign and rather counteracts the emergence of
an NGFP. Hence, diamagnetic effects work against Asymptotic Safety. Note that the culprit
is alone the diamagnetism of the graviton; the ghosts make a negative contribution to
$\eta_N^\text{total dia}$ and actually favor an NGFP.
\medskip

\noindent
\textbf{(v)} As yet we employed the optimized cutoff in this subsection. In order to
assess the degree of universality of our findings we now check them against calculations
with other cutoffs. For this purpose we choose again the one-parameter family of
exponential shape functions, $\RN_s(z) = \frac{sz}{e^{sz}-1}$, and re-compute the RG flow
for various values of the ``shape parameter'' $s$. Concerning the fixed point data, Table
\ref{tab:FPvalues} lists the remarkable result of this analysis. While the NGFP is
``destroyed'' for any cutoff when retaining the diamagnetic terms only, it persists in
all cases when only the paramagnetic contributions are kept. The fixed point coordinates
and critical exponents in the latter case are almost the same as those obtained with the
full $\eta_N$. Thus our findings appear to be perfectly stable under changes of the
cutoff: paramagnetic dominance is not a peculiarity of the optimized shape function.

We can conclude that \emph{the formation of an NGFP in the RG flow of QEG is a universal
result of the paramagnetic interaction of the metric fluctuations with their background}.

\begin{table}
\centering
\begin{tabular}{crlrlrlrl}
	\hline
	Cutoff			&	$g_*$	&	$g_*^\text{para}$	&	$\lambda_*$	&	$\lambda_*^\text{para}$	&
		$\theta_1$	&	$\theta_1^\text{para}$	&	$\theta_2$	&	$\theta_2^\text{para}$	\\
	\hline
	\hline
	$\;\;\, s=0.5$	&	\quad$0.168$	&	$0.167$	&	\quad$0.448$	&	$0.429$	&	\quad$1.122$
	& $1.331$ &	\quad$6.176$	&	$5.237$	\\
	$s=1$			&	$0.272$	&	$0.271$	&	$0.359$	&	$0.333$	&	$1.420$	&	$1.536$	&	$4.327$	&	$3.869$	\\
	$s=2$			&	$0.439$	&	$0.436$	&	$0.261$	&	$0.237$	&	$1.483$	&	$1.588$	&	$3.558$	&	$3.288$	\\
	$s=5$			&	$0.835$	&	$0.828$	&	$0.154$	&	$0.138$	&	$1.530$	&	$1.605$	&	$3.020$	&	$2.976$	\\
	optimized	&	$0.707$	&	$0.707$	&	$0.193$	&	$0.192$	&	$1.475$	&	$1.255$	&	$3.043$	&	$2.712$	\\
	\hline
\end{tabular}
\caption{Fixed point data from the full $\beta$-functions compared to those taking into
	account only the total paramagnetic contribution to $\eta_N$. The numerical evaluation
	is done	for five different cutoffs: the exponential cutoff $\RN_s$ with $s=0.5, 1,
	2, 5$, and the optimized cutoff. The ``para'' results are in remarkable accordance with
	the exact ones.}
\label{tab:FPvalues}
\end{table}

\subsection[What is special in \texorpdfstring{$d=3$}{d = 3}, and what is not]
	{What is special in \bm{$d=3$}, and what is not}
\label{sec:FPQEG3d}
If one plots the Einstein-Hilbert phase portrait in $d=3$ the result looks almost the
same as the 4D diagram in figure \ref{fig:QEGfull}. One finds strong renormalization
effects and in particular a non-Gaussian fixed point. Sometimes it is -- erroneously --
claimed that this contradicts the well known fact that classical General Relativity in
$3$ dimensions has no ``physical'' propagating degrees of freedom. Indeed, the Riemann
curvature tensor in three spacetime dimensions can be expressed in terms of the Ricci
tensor and the scalar curvature. Hence, Einstein's vacuum field equations $R_\mn=0$ tell
us that the Riemann tensor vanishes identically, $R_{\rho\mu\sigma\nu}=0$. That means
there are no gravitational waves.

This can also be seen by counting the number of independent gravitational degrees of
freedom. We start with $\frac{1}{2}d(d+1)$ unknown functions, the components of the
symmetric matrix $g_\mn$, which we try to determine from the $\frac{1}{2}d(d+1)$
algebraically independent field equations $R_\mn=0$. Those, however, are not all
independent but subject to $d$ (differential!) constraints due to the Bianchi identities.
Furthermore, we must impose $d$ coordinate conditions, leading to a total of
\be
\frac{1}{2}\, d(d+1)-d-d = \frac{1}{2}\, d(d-3)
\label{eqn:dof}
\ee
independent functions which characterize a vacuum solution $g_\mn(x)$ and whose time
evolution can be inferred from Einstein's equation. In $d=4$ this fits with the 2
polarization states of a massless spin-2 particle, and in $d=3$ with the absence of
gravitational waves.

These remarks should make it clear that the number (\ref{eqn:dof}) relates to the
equations of motion. The FRGE on the other hand is a typical \emph{off-shell}
construction: no special metric plays a distinguished r\^{o}le, in particular not the
solutions of any (classical, effective, etc.) field equation; the metric is an
argument of $\Gamma_k[g_\mn,\bar{g}_\mn]$ that can be varied freely.

Thus, in particular the curvature tensor under the functional trace on the RHS of the
FRGE solely depends on the chosen argument of $\Gamma_k$ but it does not care about
$R_{\mn\rho\sigma}=0$ being a consequence of the classical (!) field equations in
$d=3$.

As for understanding the renormalization effects actually taking place in 3 dimensions
recall that the inverse fluctuation propagator is given by (\ref{eqn:GravOperator}) and
(\ref{eqn:UGrav}): $-\bar{K}^\mn_{~~\rho\sigma} \, \bar{D}^2 + \bar{U}^\mn_{~~\rs}$. This
is a nonminimal operator since $\bar{U}^\mn_{~~\rs} \neq 0$ when $\bar{R}_{\mn\rho\sigma}
\neq 0$ as it is the case when we project on $\int\sg\, R$. Being off-shell, we allow for
paramagnetic terms now. As a result, the RG running becomes nontrivial. Reconsidering
eq.\ (\ref{eqn:etaNtot}),
\be
\eta_N =\, \frac{1}{3\Gamma(d/2)}\, (4\pi)^{1-\frac{d}{2}} \Big[ \,
	\big\{ d(d-3) \big\}_\text{total dia} + \big\{ -12(d-1)-48/d \big\}_\text{total para}\,
	\Big] g	+ \ord(g^2),
\ee
which was obtained by setting $\lambda=0$, we see that there is obviously no diamagnetic
contribution to $\eta_N$ for $d=3$, but it is nonzero thanks to the total paramagnetic
term. This is paramagnetic dominance in its most distinct form.

We want to point out that this property is universal: the prefactor of the diamagnetic
part is proportional to $d(d-3)$, independent of the cutoff. Without the paramagnetic
component we would indeed encounter the trivial case of a vanishing anomalous dimension
such that the dimensionful Newton constant would no longer be $k$-dependent. But here the
paramagnetic effects determine the RG behavior completely and provide for a non-Gaussian
fixed point.

For $\lambda \neq 0$ the diamagnetic contribution to $\eta_N$ does not vanish
identically due to an incomplete cancellation between gravitons and ghosts. However,
the paramagnetic dominance is still very pronounced.\footnote{At the fixed point for
instance, based on the optimized cutoff, the total paramagnetic contribution to $\eta_N$
is more than 27 times larger than the diamagnetic one.} This results in a situation
similar to the four-dimensional case. The flow induced by the $\beta$-functions with the
full $\eta_N$ shows an NGFP. Taking into account paramagnetic terms only one observes
qualitatively the same picture. In contrast, with diamagnetic interactions only, there is
no nontrivial fixed point.

\subsection[The \texorpdfstring{$\beta$}{b}-function of \texorpdfstring{$g$}{g} in
	\texorpdfstring{$2+\epsilon$}{2 + e} dimensions]{The $\bm{\beta}$-function of
	$\bm{g}$ in \bm{$2+\epsilon$} dimensions}
\label{sec:FPQEG2d}
As Newton's constant becomes dimensionless in two dimensions the RG flow of the
gravitational average action can be expected to show a certain degree of universality if
$d=2+\epsilon$ for $\epsilon$ small. This case has been studied in detail in ref.\
\cite{mr} already where it was shown that there exists an NGFP whose coordinates $g_*$
and $\lambda_*$ are of order $\epsilon$. Since $\lambda_k=\ord(\epsilon)$ near the fixed
point, the flow in its vicinity is described by RG equations in which we may expand the
threshold functions for small $\lambda_k$. For those occurring in $\eta_N$ this yields the
\emph{universal} leading order $\Phi_1^2(\lambda_k)=\Phi_1^2(0)+\ord(\epsilon)$ and
$\Phi_0^1(\lambda_k)=\Phi_0^1(0)+\ord(\epsilon)$ where $\Phi_1^2(0)=1$ and
$\Phi_0^1(0)=1$ for any cutoff shape function $\RN$. As a result, the leading order (in
$\epsilon$) contribution to the anomalous dimension reads $\eta_N=-b\, g+\ord(g^2)$ where
the coefficient $b$, in its decomposed form, follows from (\ref{eqn:etaNexpansion}):
\be
b = \frac{1}{3}\, \Big[ \{-6\}_\text{dia} + \{8\}_\text{ghost-dia}
	+ \{12\}_\text{para} + \{24\}_\text{ghost-para} \Big] \,.
\label{eqn:bcoeff}
\ee
The quantity $b$ is defined as in Weinberg's paper \cite{Weinberg}, so that the
NGFP-condition $d-2+\eta_N=\epsilon-b\, g_*=0$ leads to the fixed point coordinate
$g_*=\epsilon/b+\ord(\epsilon^2)$. According to our result (\ref{eqn:bcoeff}), or
\be
b = \frac{2}{3}\, \Big[ \{1\}_\text{total dia} + \{18\}_\text{total para} \Big]
	= \frac{38}{3} \,,
\label{eqn:bcoefftot}
\ee
the crucial coefficient $b$ is positive -- and the anomalous dimension negative therefore
-- not only thanks to the large ``para'' contribution but also because of the smaller,
but positive diamagnetic one. This is exactly as it should be since we know that below
$d=3$ the diamagnetic interaction drives $\eta_N$ in the same direction as the
paramagnetic, see figure \ref{fig:etaNtot}.

In the literature \cite{GKT,CD,Brown,Tsao,Kawai,JackJones,membrane} there has been a
considerable amount of confusion about the correct value of $b$. The situation has
already been discussed by Weinberg \cite{Weinberg} but was never resolved satisfactorily.
In \cite{Weinberg} two classes of disagreeing results for gravity coupled to various
matter fields were quoted.\footnote{For a more recent account see \cite{Hamber-book}.}
Here we list them for the case of $\ns$ dynamical scalar fields coupled to quantum
gravity. According to \cite{Tsao,Kawai,JackJones,membrane} the coefficient $b$ reads in
this case
\be
b = \frac{38}{3} - \frac{2}{3}\,\ns = \frac{2}{3}\big[ 19-\ns \big] \,.
\label{eqn:bcamp38}
\ee
The authors of refs.\ \cite{GKT,CD} find instead
\be
b = \frac{2}{3} - \frac{2}{3}\, \ns = \frac{2}{3}\big[ 1 -\ns \big] \,,
\label{eqn:bcamp2}
\ee
which has the same scalar contribution\footnote{The calculations disagree, however, on
the matter contributions from higher spin fields.} but differs in its pure gravity part.
Comparing the results of the two camps to the answer obtained by means of the effective
average action, eq.\ (\ref{eqn:bcoefftot})\footnote{Eq.\ (\ref{eqn:bcoefftot}) holds for
pure gravity. Adding $\ns$ scalar fields, see section \ref{sec:scalar-single} below, the
complete FRGE result reads $b=\frac{2}{3}\big[ 19-\ns \big]$, in perfect agreement with
(\ref{eqn:bcamp38}).},
we observe that \emph{the first candidate,
the coefficient in (\ref{eqn:bcamp38}), amounts to the full, i.e.\ dia- plus paramagnetic
gravity contribution, while the second, eq.\ (\ref{eqn:bcamp2}), consists of the
diamagnetic one only.} Looking at the details of their respective derivations one can see
that the different treatment of the paramagnetic piece is indeed the source of the
disagreement.

As pointed out by Weinberg \cite{Weinberg} the original expectation was that by virtue of
the count (\ref{eqn:dof}) the graviton contribution near $d=2$ should equal
$\frac{1}{2}d(d-3)=-1$ times the contribution of a single scalar. This would favor
(\ref{eqn:bcamp2}) over (\ref{eqn:bcamp38}), i.e.\ $b=\frac{2}{3}$ rather than
$b=\frac{38}{3}$ for pure gravity. However, as we emphasized in the previous subsection,
this notion of ``degrees of freedom'' refers to the fields' propagation characteristic
the diamagnetic interaction, but not the paramagnetic, is connected to. In this way we
can understand why in the framework of the average action and the FRGE the perhaps
counterintuitive result (\ref{eqn:bcamp38}) occurs, with a total gravitational
contribution 19 times stronger than that of a scalar.

Interestingly enough, the result (\ref{eqn:bcamp2}) found in \cite{GKT,CD} is by no means
computationally wrong, but rather amounts to a different definition of a ``running Newton
constant'', namely via the coefficient of the Gibbons-Hawking surface term. And indeed,
as we discuss next, when we use the FRGE to compute the running of this boundary Newton
constant the result we find is in perfect agreement with (\ref{eqn:bcamp2}).

\subsection{Adding a boundary term}
\label{sec:surface}
Recently \cite{boundary} a generalization of the Einstein-Hilbert truncation for
spacetime manifolds $\mathcal{M}$ with a nonempty boundary $\p\mathcal{M}$ has been
considered. In \cite{boundary} the truncation ansatz (\ref{eqn:Einstein-Hilbert}) was
augmented by a Gibbons-Hawking term \cite{GH} with a running prefactor parametrized by a
surface Newton constant $G_k^\p$:
\be
\Gamma_k^\p[g] = -\frac{1}{8\pi G_k^\p} \int_{\p\mathcal{M}} \textrm{d}^{d-1}x\,
	\sqrt{H}\, K \,.
\ee
Here $K$ denotes the trace of the extrinsic curvature of $\p\mathcal{M}$ embedded in
$\mathcal{M}$ and $H_\mn$ is the boundary metric induced by $g_\mn$. The normalization of
$\Gamma_k^\p$ is such that for $G_k=G_k^\p$ the disturbing surface terms in the
$\delta g_\mn$-variation cancel exactly. The RG equation for the dimensionless
$g_k^\p \equiv  k^{d-2} G_k^\p$ was found to be $\p_t g_k^\p = [d-2+\eta_N^\p] g_k^\p$
where the surface anomalous dimension $\eta_N^\p$, to leading order in the ``bulk''
Newton constant $g_k$, is given by
\be
\eta_N^\p(g^\p,g,\lambda) = \frac{1}{3}(4\pi)^{1-\frac{d}{2}}\left[ d(d+1)\,
	\Phi_{d/2-1}^1(-2\lambda) -4d\,\Phi_{d/2-1}^1(0) +\ord(g)\right]g^\p \,.
\label{eqn:etaboundary}
\ee
Going through the derivation of (\ref{eqn:etaboundary}) it is easy to see that in leading
order \emph{the surface anomalous dimension $\eta_N^\p$ is of entirely diamagnetic
origin}. In fact, eq.\ (\ref{eqn:etaboundary}) has exactly the same structure as the
diamagnetic terms in the corresponding ``bulk'' formula (\ref{eqn:etaNexpansion}).
Therefore, if it was not for the additional paramagnetic terms in
(\ref{eqn:etaNexpansion}) the equality $G_k = G_k^\p$ would be stable under RG evolution.
(The interested reader is referred to \cite{boundary} for further details.)

Thus we understand where the different (and, in fact, opposite in $d=4$) running of $G_k$
and $G_k^\p$ found in \cite{boundary} comes from: it is due to the paramagnetic
interaction which affects the running of $G_k$, but not of its surface counterpart
$G_k^\p$. Consistent with that, at $\lambda=0$, the surface anomalous dimension vanishes
in $d=3$, and it becomes
\be
\eta_N^\p = - b^\p\,g^\p + \cdots \qquad\quad \text{with }\, b^\p = \frac{2}{3}
\ee
in $d=2+\epsilon$.

Remarkably, the FRGE result in $2+\epsilon$ dimensions for the coefficient in the
boundary anomalous dimension, $b^\p = \frac{2}{3}$, coincides precisely with the gravity
contribution of (\ref{eqn:bcamp2}) found in \cite{GKT,CD}. This confirms that the authors
advocating (\ref{eqn:bcamp2}) actually computed the anomalous dimension of the boundary
Newton constant $G_k^\p$, while those who obtained (\ref{eqn:bcamp38}) focused on the bulk
quantity $G_k$. Thus, in a way, both ``camps'' are right, but their respective results,
$\eta_N^\p$ and $\eta_N$, are unavoidably different as a consequence of the paramagnetic
interaction.

\section{A matter induced bimetric action}
\label{sec:Bimetric}
Every gravitational field theory has to cope with the issue that the ``stage'' it is
constructed on, i.e.\ spacetime and its metric, is a priori not given. It should rather
be an outcome of the theory. One way out of this conceptual problem is to introduce a
background field \cite{dewitt-books}: we choose a fixed but arbitrary background metric
$\bar{g}_\mn$, and then base the quantization and the construction of possible actions on
this metric. The natural requirement of background covariance claims that physical
quantities should be independent of the choice of the background in the end. This is what
we meant by ``paradoxical'' in the introduction: one realizes background independence by
using background fields.

As a consequence, the effective average action
$\Gamma_k \equiv \Gamma_k[g_\mn,\bar{g}_\mn]$ depends on two metrics, the background
$\bar{g}_\mn$ and the dynamical metric $g_\mn=\bar{g}_\mn+h_\mn$. It may consist of
monomials built from $g_\mn$ like $\int \sg$ and $\int \sg\, R$, but also of their
background analogs $\int \sgb$, $\int \sgb\, \bar{R}$, etc., and of mixed terms. For
calculational simplicity, however, the older computations starting with \cite{mr} all
chose truncations which contained only the former terms, combined with a gauge fixing
and a ghost action depending on $g_\mn$ and $\bar{g}_\mn$ separately. These so-called
single metric truncations  amount to a possibly severe approximation, though. This issue
was first studied in \cite{elisa2} where both metrics were retained during the entire
calculation. Such truncations are referred to as \emph{bimetric}. It is encouraging that
so far all work done in this setting \cite{elisa2,MRS1,MRS2,boundary} provided further
evidence for the Asymptotic Safety scenario: although the number of couplings has
increased, the systems analyzed still develop a non-Gaussian fixed point, despite
significant quantitative changes in comparison to the single-metric approximation.

In this section we re-investigate the r\^{o}le of nonminimal, i.e.\ paramagnetic, terms,
now within a bimetric truncation. We choose an ansatz for a gravity+matter system similar
to the one in \cite{MRS1} so that the $\beta$-functions are easily evaluated within the
induced gravity approximation. The question we would like to answer is to what extent our
findings on paramagnetic dominance change when one goes beyond a single-metric theory
space. Does the bimetric extension modify the qualitative picture about the relative
importance of minimal and nonminimal terms?

\subsection{Nonminimally coupled scalar fields -- single-metric}
\label{sec:scalar-single}
To introduce the matter coupled model we start from its single-metric truncation.
We make an ansatz for $\Gamma_k$ involving the Einstein-Hilbert action with appropriate
gauge fixing and ghost terms as in (\ref{eqn:Einstein-Hilbert}), supplemented by an
action for $\ns$ scalar fields $A_i$ with mass $\bar{m}$, nonminimally coupled to
the spacetime curvature:
\be
\Gamma_k[g,A,\bar{g}] = \Gamma_k^\text{EH} + \Gamma_k^\text{gf}+ \Gamma_k^\text{gh}
	+ \frac{1}{2}\int \ddx \sg\, A_i \big( - D^2 + \bar{m}^2+ \xi R \big) A_i \,.
\ee
Here a sum over $i=1,\cdots,\ns$ is implied. In principle, $\bar{m}$ and $\xi$ are
$k$-dependent couplings but here we neglect their running. This is particularly suitable
for our purposes since we want $\xi$ to be a tunable parameter in order to test the
influence of the nonminimal term $\propto R\,A^2$.

The flow equations are derived along the lines of the previous section using the FRGE
(\ref{eqn:FRGE}), however, we take the ``large $\ns$ limit'', i.e.\ we assume
$\ns \rightarrow \infty$. In this limit the scalar field contributions to the
$\beta$-functions are dominant compared to pure gravity effects. Therefore, it is
sufficient to take into account the second functional derivative of $\Gamma_k$ with
respect only to the scalar fields in the FRGE, which is then given by
$\p_t\Gamma_k = \frac{1}{2}\, \Tr \big[ (\Gamma_k^{(2)}
	+ \Rk^\text{scalar})_{AA}^{-1} \; \p_t\Rk^\text{scalar} \big] \,$, with
\be
\big( \Gamma_k^{(2)} \big)_{AA} = -\bar{D}^2 + \bar{m}^2 + \xi \bar{R} \,,
\ee
where we have set $g_\mn = \bar{g}_\mn$ now. We refer to the gravity-scalar interactions
arising from $-\bar{D}^2$ as \emph{scalar-diamagnetic}, and to those coming from
$\xi\bar{R}$ as \emph{scalar-paramagnetic}. Employing heat kernel techniques again, we
can evaluate the functional trace and finally obtain the $\beta$-functions
\be
\beta_g = (d-2+\eta_N)g_k \qquad \text{and}\qquad \beta_\lambda = (\eta_N-2)\lambda_k 
	+ 2\ns\, g_k (4\pi)^{1-\frac{d}{2}}\, \Phi_{d/2}^1(m^2) \,,
\label{eqn:scalarbeta}
\ee
with the anomalous dimension
\be
\eta_N = \frac{2}{3}\, \ns (4\pi)^{1-\frac{d}{2}} \bigg[ \Big\{	\Phi_{d/2-1}^1(m^2)
	\Big\}_\text{scalar-dia} + \Big\{ -6\xi\, \Phi_{d/2}^2(m^2) \Big\}_\text{scalar-para}
	\bigg] g \,,
\label{eqn:scalareta}
\ee
involving the dimensionless mass $m^2\equiv \bar{m}^2/k^2$.

We observe that without the nonminimal term $\propto \xi\bar{R}\,A^2$ the anomalous
dimension $\eta_N$ would be of purely scalar-diamagnetic origin, and, provided that
$g>0$, $\eta_N$ could assume positive values only. Including the nonminimal part,
however, a second, negative, term contributes to $\eta_N$ which can change the overall
sign in (\ref{eqn:scalareta}) for appropriate values of $\xi$. As a result, there is a
very direct relation between the size of $\xi$ and the possibility of a nontrivial fixed
point with positive Newton constant.

In order to find a UV fixed point we set $m^2=0$ in (\ref{eqn:scalarbeta}) and
(\ref{eqn:scalareta}) since $\lim_{k\rightarrow\infty} m^2 = \lim_{k\rightarrow\infty}
\bar{m}^2/k^2 = 0$ in the truncation where $\bar{m}$ does not run. Solving the condition
$d-2+\eta_*=0$ for $g_*$, and $\beta_\lambda\, |_{\, g=g_*,\,\eta_N = \eta_*}=0$ for
$\lambda_*$ yields a fixed point at $(g_*,\lambda_*)$ with
\be
g_* = - \frac{3(d-2)(4\pi)^{\frac{d}{2}-1}}{2\,\ns \Big[ \Phi_{d/2-1}^1(0)
	- 6\,\xi \Phi_{d/2}^2(0) \Big]} \; \text{,\; and} \quad
\lambda_* = - \frac{3(d-2)\, \Phi_{d/2}^1(0)}{d\,\Big[ \Phi_{d/2-1}^1(0)
	- 6\,\xi \Phi_{d/2}^2(0) \Big]} \;.
\label{eqn:betalambdastar}
\ee

In eqs.\ (\ref{eqn:betalambdastar}) those terms in the square brackets that involve $\xi$
are scalar-paramagnetic contributions, the others are scalar-diamagnetic. We find a
non-Gaussian fixed point even if we switch off the former ($\xi=0$). But taking into
account the scalar-diamagnetic part alone, $\eta_*$ is negative only since the fixed
point value of Newton's constant is negative. This is something we would like to avoid
since the present system is meant to be a toy model for full fledged QEG (without matter)
which has $g_*>0$. Indeed, via the parameter $\xi$ we can change the sign of the fixed
point coordinates: for $\xi > \Phi_{d/2-1}^1(0) \big/ \big(6\,\Phi_{d/2}^2(0)\big)$
Newton's constant $g_*$ at the NGFP gets greater than zero. Hence, \emph{if the
scalar-paramagnetic interaction is strong enough it renders $g_*$ positive}.

\begin{figure}
\begin{minipage}{.49\columnwidth}
	\flushleft
	\includegraphics[width=\columnwidth]{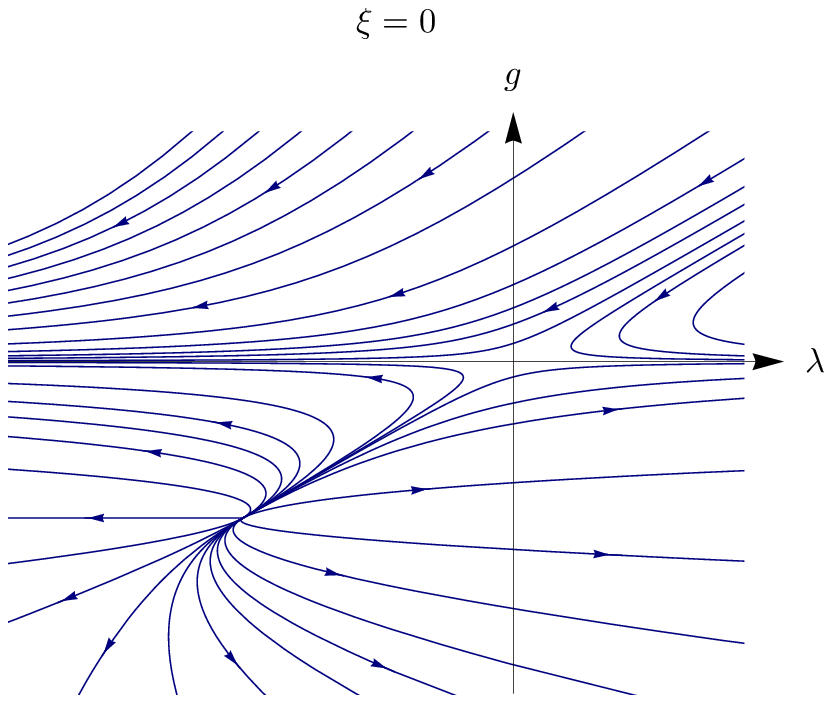}
\end{minipage}
\hfill
\begin{minipage}{.49\columnwidth}
	\flushright
	\includegraphics[width=\columnwidth]{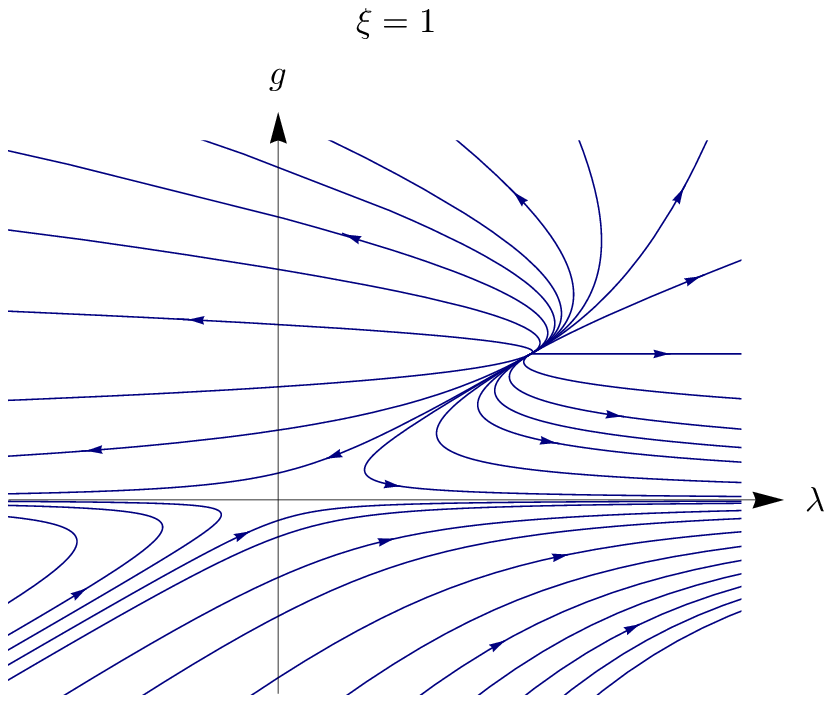}
\end{minipage}
\caption{Phase portrait for a minimal, purely scalar-diamagnetic interaction ($\xi = 0$,
	left panel) and for a nonminimal one where both scalar-dia- and scalar-paramagnetic
	terms contribute ($\xi = 1$, right panel). In the latter case we have $g_* > 0$.}
\label{fig:scalarXi}
\end{figure}

In 4 dimensions for instance, using the optimized cutoff, the sign flip of $\eta_N$
(and of $g_*$) happens at $\xi=\frac{1}{3}$. For $\xi>\frac{1}{3}$ the scalar-para term
is dominant and renders $g_*$ positive. This situation is illustrated in figure
\ref{fig:scalarXi}.

In $d=2+\epsilon$, eq.\ (\ref{eqn:scalareta}) reduces to $\eta_N = \frac{2}{3}\,\ns
\big[ 1-6\,\xi \,\big] g + \ord(\epsilon)$, for any cutoff. In the minimally coupled case
$\xi=0$ we obtain exactly the same scalar field contribution as both in eq.\
(\ref{eqn:bcamp38}) and in eq.\ (\ref{eqn:bcamp2}). For $\xi > \frac{1}{6}$, we have
$\eta_* < 0$ and $g_* > 0$ if $\epsilon$ is positive.

In the following we shall investigate if these features of the single-metric computation
survive the generalization to the bimetric truncation.

\subsection{Nonminimally coupled scalar fields -- bimetric}
\label{sec:scalar-bi}
We now perform an analysis similar to the previous subsection, but disentangle the
dynamical and the background metric. Here we follow ref.\ \cite{MRS1} and expand the pure
gravity part of the effective average action $\Gamma_k[g,A,\bar{g}] \equiv
\Gamma_k[A,\bar{h};\bar{g}]$ in terms of the metric fluctuation $\bar{h}_\mn = g_\mn -
\bar{g}_\mn$, going up to order $\ord(\bar{h}_\mn)$ only. As for the scalar fields we
generalize the truncation ansatz of \cite{MRS1} by including a nonminimal interaction
term $\propto \xi\,R A^2$:
\be
\begin{split}
\Gamma_k[A,\bar{h};\gb] = & - \frac{1}{16\pi G_k^{(0)}} \int \ddx \sgb\, \Big(
	\bar{R} - 2\Lambda_k^{(0)} \Big) \\
& + \frac{1}{16\pi G_k^{(1)}} \int \ddx \sgb\, \Big( \bar{G}^\mn-\textstyle\frac{1}{2}
	\, E_k\, \gb^\mn\bar{R} + \Lambda_k^{(1)}\gb^\mn \Big)\bar{h}_\mn \\
& + \frac{1}{2} \int \ddx \sg\; A_i\Big(  -D^2 + \bar{m}^2 + \xi\, R
	\Big) A_i \; \Big |_{g_\mn=\gb_\mn+\bar{h}_\mn} \;.
\end{split}
\label{eqn:biGammak}
\ee
In (\ref{eqn:biGammak}) $\bar{G}^\mn \equiv\, \bar{g}^{\mu\rho}\bar{g}^{\nu\sigma}\big(
\bar{R}_{\rho\sigma} - \frac{1}{2}\, \bar{g}_{\rho\sigma}\bar{R} \big)$ denotes the
Einstein tensor of the background metric. The purely gravitational part of
(\ref{eqn:biGammak}) is the most general ansatz containing no more than two derivatives,
up to first order in $\bar{h}$. The five invariants give rise to five coupling constants,
$G_k^{(0)}$, $\Lambda_k^{(0)}$, $G_k^{(1)}$, $\Lambda_k^{(1)}$, and $E_k$. Here the
superscripts $(0)$ and $(1)$ refer to the ``level'' of the corresponding parameter, i.e.\
the $\bar{h}$-order of the term in (\ref{eqn:biGammak}) in which it appears.

Again we consider the limit $\ns\rightarrow\infty$, thus we need to quantize only the
scalar fields but not the gravitons. In this case the FRGE reads
\be
\p_t\Gamma_k[A,g,\bar{g}\,] = \frac{1}{2}\, \Tr \left[ \big( 
	\Gamma_k^{(2)}[A,g,\bar{g}\,] + \Rk^\text{scalar}[\bar{g}\,] \big)_{AA}^{-1} \;
	\p_t\Rk^\text{scalar}[\bar{g}\,]\; \right] \,.
\label{eqn:FRGE-bi}
\ee
Inserting (\ref{eqn:biGammak}) into (\ref{eqn:FRGE-bi}) and evaluating the traces along
the lines of \cite{MRS1} and \cite{boundary} one finally arrives at the following system
of $\beta$-functions for the dimensionless couplings $g_k^{(0)} \equiv k^{d-2}G_k^{(0)}$,
$\lambda_k^{(0)} \equiv k^{-2} \Lambda_k^{(0)}$, $g_k^{(1)} \equiv k^{d-2} G_k^{(1)}$,
$\lambda_k^{(1)} \equiv k^{-2} \Lambda_k^{(1)}$ and $E_k$. At level $(0)$ we find:
{\allowdisplaybreaks
\begin{subequations}
\label{eqn:betalevel0}
\begin{align}
\p_t g_k^{(0)} &= \big[ d - 2 + \eta^{(0)} \big] g_k^{(0)} \,, \label{eqn:bg0}\\
\p_t\lambda_k^{(0)} &= \big[ \eta^{(0)}-2 \big] \lambda_k^{(0)} + 2\,\ns \,(4\pi)^{1
	- \frac{d}{2}}\, g_k^{(0)}\, \Phi_{d/2}^1(m^2) \,.\label{eqn:bl0}
\end{align}
\end{subequations}
Similarly one obtains for the couplings occurring at level $(1)$:
\begin{subequations}
\begin{align}
\p_t g_k^{(1)} &= \big[ d - 2 + \eta^{(1)} \big] g_k^{(1)} \,, \\
\p_t\lambda_k^{(1)} &= \big[ \eta^{(1)} - 2 \big] \lambda_k^{(1)} - \ns\,
	(4\pi)^{1-\frac{d}{2}}\, g_k^{(1)}\, \Big[(d-2)\,	\Phi_{d/2+1}^2(m^2) +
	2m^2\, \Phi_{d/2}^2(m^2)\Big] \,, \\
\p_t E_k &= \eta^{(1)} E_k + \frac{2}{3}\, \ns\, (4\pi)^{1-\frac{d}{2}}\, g_k^{(1)}
	\bigg[ \frac{d-2}{2}\,\Phi_{d/2}^2(m^2)+ 	m^2\, \Phi_{d/2-1}^2(m^2)
	\label{eqn:bE}\\* &\mkern145mu
	- 6\,\xi(d-2)\, \Phi_{d/2+1}^3(m^2) - 12\,\xi\, m^2\, \Phi_{d/2}^3(m^2)
	\, \bigg]. \nonumber
\end{align}
\end{subequations}
}As above, $m^2 \equiv \bar{m}^2/k^2$ denotes the dimensionless mass. The anomalous
dimensions at level $(0)$ and $(1)$, $\eta^{(0)}\equiv\p_t\ln G_k^{(0)}$ and
$\eta^{(1)}\equiv\p_t\ln G_k^{(1)}$, respectively, are given by
\begin{subequations}
\begin{align}
\eta^{(0)} &= \frac{2}{3}\,\ns\, (4\pi)^{1-\frac{d}{2}}\, \Big[ \big\{
	\Phi_{d/2-1}^1(m^2) \big\}_\text{scalar-dia} + \big\{ -6\,\xi\,
	\Phi_{d/2}^2(m^2) \big\}_\text{scalar-para}\, \Big]\,g_k^{(0)} \,,
	\label{eqn:eta0}\\
\eta^{(1)} &= \frac{2}{3}\, \ns\, (4\pi)^{1-\frac{d}{2}}\; \Phi_{d/2}^2(m^2)\,
	\Big[ \big\{ 1 \big\}_\text{scalar-dia} + \big\{ -6\, \xi \big\}_\text{scalar-para}\,
	\Big]\,	g_k^{(1)} \,.
	\label{eqn:eta1}
\end{align}
\end{subequations}
We observe that the system of evolution equations (\ref{eqn:bg0}) -- (\ref{eqn:eta1})
decouples. Both the level $(0)$ and the level $(1)$ couplings form closed sub-systems;
the differential equations for $g_k^{(0)}$ and $\lambda_k^{(0)}$ do not depend on
$g_k^{(1)}$, $\lambda_k^{(1)}$, $E_k$, and vice versa.

The encouraging news with regard to Asymptotic Safety is that each one of the decoupled
sets allows for both a Gaussian and a non-Gaussian fixed point. As already discussed in
detail in \cite{MRS1} the independence of the two levels entails that there exists a
total of 4 fixed points, corresponding to all possible fixed point combinations.

Moreover, having included the nonminimal term $\propto \xi\, R$ we are able to extract
additional information here. The respective scalar-diamagnetic contributions to the
anomalous dimensions at both level $(0)$ and level $(1)$ are positive for $g_k^{(0)},
g_k^{(1)} > 0$. Switching on the paramagnetic interactions, i.e.\ $\xi > 0$, however,
the values of both $\eta^{(0)}$ and $\eta^{(1)}$ decrease. For $\xi$ large enough they
become even negative. We emphasize that this behavior is not restricted to level $(0)$
but is also present at level $(1)$.

An important result of this bimetric analysis is revealed by a comparison with the
single-metric truncation: the level-$(0)$ sub-system (\ref{eqn:betalevel0}) and
(\ref{eqn:eta0}) coincides exactly with the RG equations (\ref{eqn:scalarbeta}) and
(\ref{eqn:scalareta}) one obtains in the single-metric case. Thus all conclusions drawn
in the previous subsection, in particular those concerning the fixed point, hold also for
the level-$(0)$ sub-system of the bimetric truncation. It is possible to tune the
parameter $\xi$ such that the Newton constant $g_k^{(0)}$ has a positive fixed point
value. As we will demonstrate next, the same is true for the NGFP at level $(1)$, too.

In order to search for fixed points we may set $m=0$ in eqs.\ (\ref{eqn:bg0}) --
(\ref{eqn:eta1}) as above. This results in the following NGFP coordinates of the level
$(0)$ couplings:
\be
g_*^{(0)} = - \frac{3(d-2)(4\pi)^{\frac{d}{2}-1}}{2\,\ns \left[ \Phi_{d/2-1}^1(0)
	- 6\,\xi\, \Phi_{d/2}^2(0) \right]} \; , \quad
\lambda_*^{(0)} = - \frac{3(d-2)\, \Phi_{d/2}^1(0)}{d\,\left[
	\Phi_{d/2-1}^1(0) - 6\,\xi\, \Phi_{d/2}^2(0) \right]} \;.
\label{eqn:FP0}
\ee
Likewise we find for the couplings at level $(1)$:
\be
\begin{split}
g_*^{(1)} &= - \frac{3(d-2)(4\pi)^{\frac{d}{2}-1}}{2\,\ns\,\Phi_{d/2}^2(0)\,
	\big[ 1 - 6\,\xi \big]} \;, \qquad
\lambda_*^{(1)} = \frac{3(d-2)^2\, \Phi_{d/2+1}^2(0)}{2\, d\,
	\Phi_{d/2}^2(0)\,\big[ 1 - 6\,\xi \big]} \;, \\
E_* &= - \frac{d-2}{2\,\big[ 1 - 6\,\xi \big]}\, \left[ 1 - 12\, \xi\,
	\frac{\Phi_{d/2+1}^3(0)}{\Phi_{d/2}^2(0)} \right] \,.
\end{split}
\label{eqn:FP1}
\ee

The two equations in (\ref{eqn:FP0}) clarify our statement made above:
there is a critical value $\xi_\text{crit}^{(0)} = \Phi_{d/2-1}^1(0) \big/
\big(6\,\Phi_{d/2}^2(0)\big)$ where the scalar field contribution to $g_*^{(0)}$ and
$\lambda_*^{(0)}$ changes their signs. For $\xi < \xi_\text{crit}^{(0)}$ it is the
scalar-diamagnetic part that decides about the sign of the anomalous dimension
$\eta^{(0)}$; in this case $\eta_*^{(0)}<0$ is possible only if $g_*^{(0)}$ and
$\lambda_*^{(0)}$ are negative. However, tuning $\xi$ to higher values,
$\xi > \xi_\text{crit}^{(0)}$, the scalar-paramagnetic interaction gets dominant and
flips these signs, rendering the fixed point coordinates $g_*^{(0)}$ and
$\lambda_*^{(0)}$ positive.\footnote{At the critical value $\xi = \xi_\text{crit}$ the
scalar field contribution to the anomalous dimension $\eta^{(0)}$ drops out. Then the
limit $\ns \rightarrow \infty$ is no longer admissible, and one can not neglect other
fields. In this case $g_*^{(0)}$ and $\lambda_*^{(0)}$ contain non-scalar-field
contributions which prevent them from diverging.}

Remarkably, a very similar behavior occurs at level $(1)$. Again we find a critical value
for $\xi$, $\xi_\text{crit}^{(1)} = 1/6$, above which the signs of $g_*^{(1)}$ and
$\lambda_*^{(1)}$ are flipped. In particular $g_*^{(1)}$ is rendered positive for $\xi$
large enough. As a consequence, for $\xi > \max \big(\xi_\text{crit}^{(0)},
\xi_\text{crit}^{(1)} \big) \,$ we have both $g_*^{(0)} > 0$ and $g_*^{(1)} > 0$.

The similarities between the various levels can be understood as a reflection of the
basic split symmetry \cite{MRS1}; while the cutoff breaks it to some extent, it still has
a certain impact on the RG flow.

Now we can return to the question raised above: does the transition from single- to
bimetric truncations destroy our picture of the r\^{o}le of paramagnetic terms? According
to the findings of this section the answer is no. All qualitative features contained in
the single-metric result reappear in the bimetric setting. The RG flow with its fixed
points, and the influence of the scalar-paramagnetic interaction found in the
single-metric computation, is recovered in its bimetric analog both at level $(0)$ and
level $(1)$.

Thus, at least for the matter induced truncation investigated here, there remains the
special significance of the paramagnetic effects.

\section{QEG spacetimes as a polarizable medium}
\label{sec:QEGvacuum}
The previous sections dealt with the predominance of paramagnetic interactions over
diamagnetic ones in determining certain gross features of the RG flow in QEG. As we
emphasized in section \ref{sec:vacuum} already, a priori this fact has nothing to do with
the interpretation of the quantum field theory vacuum as a paramagnetic medium.

Nevertheless, as we shall argue in this section, the spacetimes of QEG can be seen as a
polarizable medium, indeed with a ``paramagnetic'' response to external perturbations. In
this respect they are analogous to the vacuum state of Yang-Mills theory.

As a useful application of our results about paramagnetic dominance on the technical
side, we explain in this section also the emergence of decoupling scales due to the
nonminimal character of the kinetic operator. They are important for ``RG-improving''
lower order calculations, as we shall illustrate by means of a simple black hole example.

\subsection{Physical vs.\ cutoff scales}
\label{sec:PhysVsCutoff}

\subsubsection{Decoupling scales and RG improvement}
\label{sec:decoupling}
By definition, we say that the effective average action $\Gamma_k$ displays (complete)
\emph{decoupling} if, when we lower the IR cutoff $k$, it stops running at a certain
finite scale $k = k_\text{dec} > 0$. When this happens the remainder of the RG evolution
from $k_\text{dec}$ down to $k=0$ is ``for free'', and the ordinary effective action is
$\Gamma \equiv \Gamma_0 = \Gamma_{k_\text{dec}}$. Generically there will be only partial
decoupling, i.e.\ only the contributions to the $\beta$-functions of particular fields,
or modes, vanish at $k_\text{dec}$, or only some terms in $\Gamma_k$ might stop running.
Hence, there is more than one decoupling scale in general.

In its most general form, decoupling happens whenever certain terms in the Hessian
$\Gamma_k^{(2)}$ dominate over the cutoff operator $\Rk$. In this case we have, roughly
speaking, ``$\Gamma_k^{(2)} \gg \Rk$'', and also ``$\Gamma_k^{(2)} \gg \p_t \Rk$'' since
both $\Rk \sim k^2$ and $\p_t \Rk \sim k^2$. Then it is clear that the RHS of the FRGE,
$\p_t \Gamma_k = \frac{1}{2}\, \Tr \big[ (\Gamma_k^{(2)} + \Rk)^{-1}\, \p_t\Rk \big]$,
becomes very small such that $\Gamma_k$ no longer runs significantly with $k$. Usually it
is not easy to find out when this happens since $\Gamma_k^{(2)}$ and $\Rk$ are
non-commuting operators in general, and a certain understanding of the spectrum of
$\Gamma_k^{(2)} + \Rk$ is necessary.

Nevertheless, the FRGE reproduces of course the simple examples known from perturbation
theory. If, for instance, $\Gamma_k$ contains some mass term freezing out at
$k = \bar{m}_0$ we have for the IR modes, symbolically, $\Gamma_k^{(2)}+\Rk = \cdots +
\bar{m}_0^2 + k^2 + \cdots$ so that $\Gamma_k \approx {}$const for $k \lesssim\bar{m}_0$.

The situation becomes more interesting when the decoupling scale is field dependent.
Consider for instance the $\phi^4$-truncation for a single scalar field with
$\Gamma_k^{(2)} = -\Box + \bar{m}_k^2 + \lambda_k\phi^2$. In the massless regime
($\bar{m}_k \ll k$) decoupling occurs when $k^2$ gets negligible relative to $\lambda_k
\phi^2$. Thus $k_\text{dec} \equiv k_\text{dec}(\phi) = \lambda_k^{1/2}\phi$. Knowing
this decoupling threshold we can predict terms in $\Gamma_{k\rightarrow 0}$ which were
not included in the truncation used to determine the running of the couplings
$\bar{m}_k$, $\lambda_k$, $\cdots$. In fact, a simple $\phi^2 + \phi^4$ ansatz is
sufficient to find the well known logarithmic running of the quartic coupling: $\lambda_k
\propto \ln k$. Thus decoupling predicts that the potential in $\Gamma = \Gamma_{k=0}$
contains a term $\lambda_{k_\text{dec}} \phi^4 = \ln \big(k_\text{dec}(\phi)\big)\phi^4$
which equals $\phi^4 \ln \phi$ to leading order for strong fields $\phi$. This term is
in fact exactly the correct Coleman-Weinberg potential of a massless scalar theory.

Obviously the impressive power of this decoupling argument \cite{wet-dec} resides in the
fact that it sometimes allows the computation of terms in the effective action
\emph{which were not included in the truncation ansatz}.
(See \cite{livrev} and \cite{h1} for a more detailed discussion.)

In the $\int (F_\mn^a)^2$-truncation of Yang-Mills theory where $\Gamma_k^{(2)} = -D^2
+ 2i g_k\, F$ it can happen that the nonminimal $F$-term dominates over $k^2$, suggesting
a decoupling scale which depends on the field strength, $k_\text{dec}^4(F) \approx
g_{k_\text{dec}}^2\, F_\mn^a F^{a\,\mn}$, implying that $\ln \big(k_\text{dec}^4(F)\big)
= \ln(F_\mn^a F^{a\,\mn})$, again omitting a field independent constant which is
subdominant for strong fields. This insight into the decoupling features of $\Gamma_k$
allows us to predict an $F^2 \ln(F^2)$-term in $\Gamma_0$ from the knowledge of
$\beta_{g^2}$ in the $F^2$-truncation alone: $\Gamma[F] = \Gamma_{k_\text{dec}(F)}[F]
= \frac{1}{4}\int g_{k_\text{dec}(F)}^{-2}(F_\mn^a)^2$.

Taking the generic model with the $\beta$-function (\ref{eqn:beta0diapara}) as an
example, the $k$-dependence of the gauge coupling is given by $g_k^{-2} = g_\Lambda^{-2}
\big[ 1 - \frac{1}{2} g_\Lambda^2\, \beta_0 \ln(k^2/\Lambda^2) \big]$ with an arbitrary
constant of integration $\Lambda$. This leads to $\Gamma[F] = \int\textrm{d}^4 x\,
\mathcal{L}_\text{eff}$ with the effective Lagrangian
\be
\mathcal{L}_\text{eff} = \frac{1}{4 g_\Lambda^2}\, F_\mn^a F^{a\,\mn} \left[ 1 -
	\frac{1}{4}\, g_\Lambda^2\, \beta_0 \ln\big(F_\mn^a F^{a\,\mn}/\Lambda^4\big)\right]\,.
\label{eqn:YMLeff}
\ee
It can be checked by conventional means \cite{dittrich} that (\ref{eqn:YMLeff}) coincides
with the correct one-loop effective action in the strong field limit, meaning that terms
$\propto DDF$ are neglected relative to $F^2$-terms.

As an application we mention that (upon scaling $g_\Lambda$ into the gauge field and Wick
rotating) the effective Lagrangian (\ref{eqn:YMLeff}), by virtue of
$- \frac{\p\mathcal{L}_\text{eff}}{\p B} = H = (1+\chi_\text{mag})^{-1}B \approx B -
\chi_\text{mag}\, B$, yields the magnetic susceptibility (\ref{eqn:chiYM}) which we
discussed earlier. Here we see that it is valid within logarithmic accuracy, when field
gradients can be neglected relative to field amplitudes.

For Lorentzian signature, the RG improvement consists in replacing
$\frac{1}{2g_\Lambda^2} \big(\mathbf{E}^2 - \mathbf{B}^2\big)$ with
\be
\frac{1}{2 g_k^2} \Big(\mathbf{E}^2 - \mathbf{B}^2\Big) \equiv \frac{1}{2 g_\Lambda^2}
\left( \varepsilon_k \mathbf{E}^2 - \frac{1}{\mu_k}\mathbf{B}^2 \right) \,,
\label{eqn:LeffNew}
\ee
which is to be evaluated at $k = k_\text{dec}(F)$. This amounts to defining the scale, or
field dependent dielectric constant and the magnetic permeability by
\be
\varepsilon_k = \frac{1}{\mu_k} = \frac{g_\Lambda^2}{g_k^2} \;.
\label{eqn:defpermeability}
\ee
Hence, $\varepsilon_k \mu_k = 1$ is satisfied automatically. The normalization is fixed
such that $\varepsilon_\Lambda = \mu_\Lambda = 1$ at the UV scale $\Lambda$, and when the
deviations of $\varepsilon$ and $\mu$ are small we have approximately
$\chi_\text{el} = -\chi_\text{mag}$.

The latter relation allows us to deduce the $E$-dependence of $\varepsilon(E) = 1 +
\chi_\text{el}(E)$ from the $\chi_\text{mag}(B)$ in eq.\ (\ref{eqn:chiYM}). For later
comparison with the gravitational case we note that the (real part of the) corresponding
effective Lagrangian in a static (color-) electric field $\mathbf{E} \equiv -
\bm{\nabla}\Phi_\text{el}$ reads
\be
\mathcal{L}_\text{eff} = \frac{1}{2}\, \varepsilon\big(|\bm{\nabla}\Phi_\text{el}|\big)
	|\bm{\nabla}\Phi_\text{el}|^2 \;,\qquad \varepsilon\big(|\bm{\nabla}\Phi_\text{el}|\big)
	= 1 + \frac{1}{2}\, \beta_0\, g_\Lambda^2 \,
	\ln\left(\frac{\Lambda^2}{g_\Lambda |\bm{\nabla} \Phi_\text{el}|} \right).
\ee
We see that for $\beta_0>0$ ($\beta_0<0$), in QED (QCD), say, integrating out fluctuation
modes between the effective IR cutoff $k^2 = g_\Lambda |\bm{\nabla} \Phi_\text{el}|$ and
$\Lambda^2$ leads to a dielectric constant $\varepsilon>1$ ($\varepsilon<1$) indicating
that test charges are screened (antiscreened) by the virtual excitations populating the
vacuum \cite{DR1}.

\subsubsection{Decoupling scales in QEG}
\label{sec:decouplingQEG}
In the Einstein-Hilbert approximation of QEG the inverse propagator of the metric
fluctuations $h_\mn$ is of the form (\ref{eqn:GravOperator}), i.e.\ $-\bar{K}\bar{D}^2
+ \bar{U}$. The tensor $\bar{U}$ was given in eq.\ (\ref{eqn:UGrav}). In this subsection
we restrict ourselves to $\Lambda_k=0$ so that $\bar{U}^\mn_{~~\rho\sigma} = 0$ when
the Riemann tensor vanishes. For suitable backgrounds $\bar{g}_\mn$ we can distinguish
two forms of decoupling in which $k^2$ can be neglected relative to either $-\bar{K}
\bar{D}^2$ or $\bar{U}$, respectively.

\noindent
\textbf{(A)} The first case is realized in flat spacetimes ($\bar{U} = 0$, $\bar{D}^2 =
\p^2$), with a finite volume, say, where $-\p^2$ has a lowest eigenvalue $p_\text{min}^2$
and so decoupling occurs at $k_\text{dec} = p_\text{min}$. More generally, when the
background is not too strongly curved, the usual intuition about Fourier analysis still
applies and geometrical constraints involving (proper) length scales $L$ affect (i.e.,
cut off) the spectrum of $-\bar{D}^2$ near $1/L^2$. For example, at large distances from
a Schwarzschild black hole the spacetime curvature is weak ($\bar{U} \approx 0$) and only
the Laplacian $-\bar{D}^2$ matters. If we focus on the portion of spacetime interior to a
sphere of constant Schwarzschild radial coordinate $r$, decoupling should occur near
\be
k_\text{dec}^{(\text{A})}(r) \approx \frac{1}{r} \;.
\label{eqn:kdecA}
\ee

While there exists no rigorous proof, a number of strong arguments in favor of
(\ref{eqn:kdecA}) has been put forward in the literature, and its consequences have been
studied \cite{mr,bh,h1}. In particular $k(r)\propto 1/r$ is the only function $k=k(r)$
one can write down which respects spherical symmetry and introduces no new scale into
the problem. Moreover, in the analogous problems of standard quantum field theory (QED,
QCD, etc.), eq.\ (\ref{eqn:kdecA}) is known to be the correct cutoff. In QED, for
instance, the RG improvement of the Coulomb potential on the basis of (\ref{eqn:kdecA})
does indeed lead to the correct Uehling-Serber potential originally determined by
standard methods. (See \cite{DR1} for a detailed discussion). In fact, this simple way
of obtaining the screened Coulomb potential works only in \emph{massless} QED. If the
electrons have a nonzero mass, a second length scale besides $r$ makes its appearance,
namely the Compton wavelength $1/m$. As a consequence, any ansatz
$k(r)=\frac{1}{r}\,f(mr)$ is consistent with the general principles, and the function
$f(mr)$ could not be guessed without an explicit (standard) calculation.

In this context, the operator $\Rk$ and the ``cutoff identification'' $k=k(r)$, or, more
generally $k=k(x^\mu)$, can be interpreted as a mathematical model of the true, physical
mode suppression process. Here we are not looking for universal features but, on the
contrary, those of a very particular physical situation.

With a cutoff identification $k=k(x^\mu)$ the scale $k$ becomes effectively a field over
spacetime. In a sense, $\Gamma_k$ with position dependent running couplings can be
thought of as an adiabatic approximation meaning that the ``fast dynamics'' of the high
frequency quantum modes has been integrated out in the background of the ``slow'' ones.
If we then set $k=k(x^\mu)$ the ``slow'' background for the ``fast'' modes varies from
point to point. This procedure can be meaningful if the $x$-dependence of $k$ is
sufficiently weak in comparison to the corresponding variation of the modes integrated
out.

\noindent
\textbf{(B)} The second case, complete or partial decoupling via the nonminimal term,
occurs when some or all eigenvalues of $\mU$ become larger than $k^2$. Hereby $\mU \equiv
\big( \bar{U}^\mn_{~~\rho\sigma} \big)$ is regarded a $\frac{d(d+1)}{2}\times
\frac{d(d+1)}{2}$ matrix whose rows and columns are labeled by the symmetric pairs
($\mn$) and ($\rho\sigma$).

For a generic background $\bar{g}_\mn$ it will be difficult in general to find the
eigenvalues of $\bar{U}^\mn_{~~\rho\sigma}$ explicitly. However, to deduce a scale of at
least partial decoupling knowledge of easily computable curvature invariants is
sufficient sometimes. For example, for Ricci flat backgrounds, i.e.\ solutions to the
classical vacuum equation $\bar{R}_\mn = 0$, we have $\bar{U}^\mn_{~~\rho\sigma} =
- \frac{1}{2}\left[ \bar{R}^{\nu~\mu}_{~\rho~\sigma}
+ \bar{R}^{\nu~\mu}_{~\sigma~\rho}\right]$, implying that $\mU$ is traceless,
$\tr(\mU) \equiv \bar{U}^\mn_{~~\mn} = 0$, and that
\be
\tr(\mU^2) = \frac{3}{4}\,\bar{R}_{\mn\rho\sigma} \bar{R}^{\mn\rho\sigma} \,.
\ee
So we can conclude that when $k^4$ drops below $k_\text{dec}^4 = \tr(\mU^2)$ there is at
least one eigenvalue of $\mU$ which has a magnitude larger than $k^2$.

\subsubsection{RG improved black holes}
\label{sec:blackHoles}
As a simple illustration, consider a Schwarzschild black hole of mass $M$. In this case
the quadratic curvature invariant is
\be
\bar{R}_{\mn\rho\sigma} \bar{R}^{\mn\rho\sigma} = 12 \left(\frac{2GM}{r^3}\right)^2 \,.
\ee
By the above argument it suggests the decoupling scale
\be
k_\text{dec}^{(\text{B})}(r) \approx \left(\frac{2GM}{r^3}\right)^{1/2}
	\equiv \frac{1}{r}\, \sqrt{\frac{2GM}{r}} \;,
\ee
where we have discarded a factor of order unity.

By now we have found two candidates for position dependent decoupling scales in the
Schwarzschild spacetime, namely $k_\text{dec}^{(\text{A})} \propto r^{-1}$ and
$k_\text{dec}^{(\text{B})} \propto r^{-3/2}$ which are related to the fluctuations'
diamagnetic ($\bar{D}^2$) and paramagnetic ($\bar{U}$) interaction with the background,
respectively. With regard to the RG improvement of the black hole spacetime proposed in
\cite{bh,erick} it is interesting to ask which one is more effective, i.e.\ is located
at the higher scale. The answer is seen to depend on whether $r$ is smaller or larger
than the pertinent Schwarzschild radius $r_\text{S} \equiv 2GM$:
\be
k_\text{dec}^\text{max}(r) = \left\{
\begin{array}{ll}
\quad \displaystyle\frac{1}{r}&\text{for}\quad r > r_\text{S} \qquad
\text{(diamagnetic interaction)} \\[2mm]
\displaystyle\frac{1}{r}\,\sqrt{\frac{r_\text{S}}{r}}\quad &\text{for}\quad r < r_\text{S}
\qquad \text{(paramagnetic interaction)}
\end{array}
\right.
\label{eqn:BHdecoupling}
\ee
It is quite intriguing that approaching a black hole it happens already at the horizon
scale (rather than, say, the Planck scale) that the paramagnetic effects take over; for
a heavy black hole this is a perfectly macroscopic scale, after all.\footnote{Of course,
this is not to say that there are necessarily large quantum effects at the horizon; this
requires a significant RG running of the couplings near $k_\text{dec}$ which (according
to the present truncations) occurs near the Planck scale only.}

Remarkably enough, (\ref{eqn:BHdecoupling}) coincides exactly with the cutoff
identification motivated in refs.\ \cite{bh} by means of an entirely different
reasoning.\footnote{In particular the typical $k \propto r^{-3/2}$ behavior at short
distances was found in \cite{bh}. It turned out that, for large distances, it
\emph{must} get replaced by $k \propto r^{-1}$ since otherwise Donoghue's \cite{don}
perturbative correction to Newton's potential would not be recovered. From the present
point of view this implies that, first, the perturbative result \cite{don} is of a
``diamagnetic'' nature and, second, RG-improvement based upon the cutoff identification
$k_\text{dec}^4 \propto \bar{R}_{\mn\rho\sigma}^2$ is demonstrably wrong for $r \gg
r_\text{S}$ as it contradicts explicit perturbative computations.} In \cite{bh} it was
used to explore the leading QEG corrections to the Schwarzschild metric, in particular
the modified horizon and causal structure and the related quantum corrected
thermodynamics were analyzed. Indications were found that the Hawking evaporation process
might come to a halt when $M \approx m_\text{Pl}$, and that the central singularity
either disappears completely or at least is significantly ameliorated by the quantum
effects. In retrospect we now understand that these effects are all predominantly due to
the paramagnetism of the metric fluctuations.

\subsubsection{Gravitational effective Lagrangian by RG improvement?}
\label{sec:EffLagrangian}
Let us now try to follow the discussion in section \ref{sec:decoupling} as closely as
possible for gravity. As far as a simple derivation of $\Gamma \equiv \Gamma_{k=0} =
\int \text{d}^4 x\, \mathcal{L}_\text{eff}$ in the strong field strength/curvature regime
is concerned, the method is less powerful in QEG than in Yang-Mills theory, however,
the reason being that there is more than one invariant that could take over the r\^{o}le
of $F_\mn^a F^{a\,\mn}$ in determining the decoupling scale. For the regime of strong,
constant field strength (curvature) a general quadratic ansatz reads
\be
k_\text{dec}^4 = c_1 R_{\mn\rho\sigma}^2 + c_2 R_\mn^2 + c_3 R^2 \,,
\label{eqn:kdec}
\ee
with dimensionless constants $c_i$. In order to evaluate $\Gamma_k = \frac{1}{16\pi G_k}
\int\text{d}^4 x \sg\,(-R + 2 \Lambda_k)$ at $k = k_\text{dec}$ let us use the following
simple but qualitatively correct caricature of the RG trajectories of the Einstein-Hilbert
truncation \cite{bh}:
\be
\frac{1}{G_k} = \frac{1}{G_0} + \frac{k^2}{g_*} \;, \qquad\qquad
\frac{\Lambda_k}{G_k} = \frac{\Lambda_0}{G_0} + \frac{\lambda_* k^4}{g_*} \;.
\label{eqn:simpleTraj}
\ee
For fields strong enough so that $k$ is in the asymptotic scaling regime the cosmological
constant term in $\Gamma_{k_\text{dec}}$ gives rise, in $\mathcal{L}_\text{eff}$, to the
three (curvature)${}^2$ terms, precisely in the linear combination of (\ref{eqn:kdec}).
The improvement of the $\sg\,R$ term, taken literally, leads to a structure with a square
root: $R\, [ c_1 R_{\mn\rho\sigma}^2 + c_2 R_\mn^2 + c_3 R^2 \,]^{1/2}$. One can
speculate that the (as yet unavailable) exact result for $\mathcal{L}_\text{eff}$ will
amount to $c_1 = c_2 = 0$ which avoids the somewhat implausible nonlocality due to the
square root. If so, we may conclude that for gravitational fields satisfying
$R^2 \gg DDR$ and, to be in the regime where $1/G_0$ and $\Lambda_0/G_0$ in
(\ref{eqn:simpleTraj}) can be neglected, $k_\text{dec} \gg m_\text{Pl}$, the effective
Lagrangian has the structure
\be
\mathcal{L}_\text{eff} = a_1 R_{\mn\rho\sigma}^2 + a_2 R_\mn^2 + a_3 R^2 \,.
\ee
In any case, it seems fairly certain that as a consequence of paramagnetic decoupling
$\mathcal{L}_\text{eff}$ is of the (curvature)${}^2$ form; a priori this includes also
exotic dimension 4 terms like $R(R_\mn R^\mn)^{1/2}$, say \cite{alfio}.

\subsection{The QEG vacuum as a polarizable medium}
\label{sec:QEGmedium}
We stressed repeatedly the conceptual difference between the dominance of the
paramagnetic interactions and a possibly paramagnetic response to external fields. The
latter is a property of the vacuum or any other state of the quantum field theory under
consideration. While the previous sections all dealt with the dominance of the
paramagnetic interaction term over the diamagnetic one, we now turn to the question of
how the QEG vacuum responds to external fields.

We shall not embark here on a discussion of the tensorial susceptibilities one can define
for a general gravitational field but rather restrict the expectation value of the metric
to the lowest post-Newtonian order. This will display the analogy of QEG to QED or
Yang-Mills theory most clearly. Employing Cartesian coordinates $x^\mu = (t,\mathbf{x})$,
we consider metrics of the form \cite{weinberg-book}
\be
g_\mn dx^\mu dx^\nu = -(1 + 2\,\Phi_\text{grav})\, dt^2 +2\,\bm{\zeta}\cdot d\mathbf{x}\,
	dt + (1 - 2\,\Phi_\text{grav})\, d\mathbf{x}^2 \,,
\label{eqn:metricpostN}
\ee
where $\Phi_\text{grav}$ and $\bm{\zeta}$ are the gravitational scalar and vector
potentials, respectively. We assume them time independent, and adopt the harmonic
coordinate condition, $\bm{\nabla}\cdot\bm{\zeta} = 0$.

Leaving the cosmological constant aside, we consider the Lorentzian version
\cite{Lor-EAA} of the effective average action $\Gamma_k^\text{Lor}[g] =
\frac{1}{16\pi G_k} \int \text{d}^4 x \sqrt{-g}\, R[g]$. Inserting the metric
(\ref{eqn:metricpostN}) and retaining at most quadratic terms in $\Phi_\text{grav}$ and
$\bm{\zeta}$ we find
\be
\Gamma_k^\text{Lor}[g] = - \frac{1}{4\pi} \int \text{d}^4 x\,\frac{1}{2G_k}\,
	\big( \mathbf{g}^2 - \bm{\Omega}^2 \big) \,.
\label{eqn:GammaLor}
\ee
Here we encounter the acceleration $\mathbf{g} \equiv - \bm{\nabla}\Phi_\text{grav}$ and
the angular velocity of the local inertial frames, $\bm{\Omega} \equiv - \frac{1}{2}
\bm{\nabla} \times \bm{\zeta}$. They are the gravitational analogs of the
electromagnetic, or Yang-Mills, $\mathbf{E}$ and $\mathbf{B}$ fields, respectively.
Following the logic that had led us to eq.\ (\ref{eqn:defpermeability}) we now rewrite
eq.\ (\ref{eqn:GammaLor}) in the following fashion:
\be
\Gamma_k^\text{Lor}[g] = -\frac{1}{4\pi} \int \text{d}^4 x\, \frac{1}{2G_\Lambda}
	\left( \varepsilon_k^\text{grav} \mathbf{g}^2 - \frac{1}{\mu_k^\text{grav}}\,
	\bm{\Omega}^2 \right) \,.
\label{eqn:GammaLorNew}
\ee
Here $G_\Lambda$ is a scale independent prefactor, Newton's constant at some fixed UV
scale $k = \Lambda$. In this reinterpretation of the running action its $k$-dependence is
carried by the ``gravi-dielectric constant'' $\varepsilon_k^\text{grav}$ and the
``gravimagnetic permeability'' $\mu_k^\text{grav}$, defined by
\be
\varepsilon_k^\text{grav} = \frac{1}{\mu_k^\text{grav}} = \frac{G_\Lambda}{G_k} \;.
\ee
For the simplified trajectory (\ref{eqn:simpleTraj}), for example, we obtain the explicit
formula
\be
\varepsilon_k^\text{grav} = \frac{1}{\mu_k^\text{grav}} = \frac{g_* + G_0 k^2}{g_* +
	G_0 \Lambda^2} \;.
\ee
As $\frac{1}{2}(\mathbf{g}^2-\bm{\Omega}^2)$ corresponds to the Maxwell Lagrangian
$\frac{1}{2}(\mathbf{E}^2-\mathbf{B}^2)$, the analogy between (\ref{eqn:GammaLorNew}) and
its gauge theory counterpart (\ref{eqn:LeffNew}) is indeed striking.

In order to understand the physics contents of (\ref{eqn:GammaLorNew}) it is not
necessary to actually identify $k \equiv k_\text{dec}(R)$. Let us continue to consider
$\varepsilon_k^\text{grav}$ and $\mu_k^\text{grav}$ functions of the IR cutoff $k$
itself, and let us ask how they evolve along an RG trajectory. Starting at the UV cutoff
$k = \Lambda$ we have $\varepsilon_\Lambda^\text{grav} = \mu_\Lambda^\text{grav} = 1$
initially. Then, lowering $k$, the RG flow is such that $G_k$ is the larger the smaller
is $k$. Hence we see that integrating out the metric fluctuations in the momentum
interval $[k,\Lambda]$ gives rise to a gravi-dielectric constant (gravimagnetic
permeability) smaller (larger) than unity:
\be
\varepsilon_k^\text{grav} \leq 1 \,, \quad \mu_k^\text{grav} \geq 1 \qquad \text{for }\,
	k \leq \Lambda.
\ee
In this sense, \emph{the behavior of the QEG vacuum is analogous to that of Yang-Mills
theory: $\varepsilon_k^\mathrm{grav} < 1$ implies that external charges (masses) are
antiscreened, and $\mu_k^\mathrm{grav} > 1$ indicates the paramagnetic response to
external gravimagnetic fields}.

We emphasize that the RG trajectories are unaffected by the post-Newtonian approximation
used here for purely illustrative purposes. Inserting a more general argument into
$\Gamma_k^\text{Lor}[g]$, the function $G_k$ will remain the same, only the
interpretation in intuitive terms might become more difficult.

\section{Summary and conclusion}
\label{sec:conclusion}
In a large class of well understood physical systems the pertinent quantum fluctuations
$\bp$ are governed by inverse propagators of the general form $-D_\cA^2 + \mU(F_\cA)$
where $D_\cA$ is the covariant derivative with respect to a certain connection $\cA$, and
$\mU$ denotes a matrix-valued potential depending on its curvature, $F_\cA$. The first
and the second term of the quadratic Lagrangian $\mathcal{L} = \frac{1}{2}\,\bp \big(
-D_\cA^2 + \mU(F_\cA) \big) \bp$ give rise to, respectively, diamagnetic-type and
paramagnetic-type interactions of the $\bp$'s with the background constituted by the
$\cA$ field. In the regime of the interest, the two types of interactions have an
antagonistic effect, but as the paramagnetic ones are much stronger than their
diamagnetic opponents they win and thus determine the qualitative properties of the
system.

Well known examples of this ``paramagnetic dominance'' include the susceptibility of
magnetic systems, the screening of electric charges in QED, and the antiscreening of
color charges in Yang-Mills theory.

In this paper we showed that also 4-dimensional Quantum Einstein Gravity belongs to this
class of systems. The RG flow of QEG is driven by the quantum fluctuations of the metric
which, too, have an inverse propagator with the above structure. We disentangled the dia-
from the para-type contributions to the RG flow, in particular to the anomalous dimension
of Newton's constant, $\eta_N$. The negative sign of $\eta_N$ which is crucial for
gravitational antiscreening and Asymptotic Safety was found to be due to the
predominantly paramagnetic interaction of the gravitons with external gravitational
fields. Those interactions are sufficient by themselves to trigger the formation of a
non-Gaussian RG fixed point. On the other hand, the diamagnetic interaction would not
lead to such a fixed point on its own, and, in fact, in $d > 3$ dimensions it counteracts
gravitational antiscreening and Asymptotic Safety. Thus \emph{the NGFP owes its existence
to the paramagnetic dominance}.

In the familiar quantum field theories, such as QED and QCD, one of the most interesting
tasks, which often is also essential from the practical point of view, consists in
determining the properties of its vacuum state, e.g.\ the response of quantum
fluctuations to external fields. In the case of quantum gravity we were led to the
following intuitive picture of a QEG ``vacuum'' state, a spacetime represented by a
self-consistent solution $\bar{g}_\mn$ to the effective field equations for instance.

The dominant paramagnetic coupling of the metric fluctuations $h_\mn$ to their
``condensate'', that is, the background $\bar{g}_\mn$, has the form $\int h(x) \bar{U}(x)
h(x)$ which is analogous to $\int \bar{\psi} (\bm{\sigma} \cdot \mathbf{B}) \psi$ for
magnetic systems. It contains no derivatives of $h_\mn$, i.e.\ it is \emph{ultralocal},
and the interaction energy it gives rise to depends only on the spin orientation of
$h(x)$ relative to $\bar{U}(x)$ at each spacetime point $x$ individually. So the
essential physical effects in the fixed point regime are due to \emph{fluctuations which
do not correlate different spacetime points}. To the extent the orbital motion effects
caused by $\int h \bar{D}^2 h$ can be neglected, different spacetime points decouple
completely.

Thus, if one wants to invoke a magnetic analogy again, the QEG vacuum should be
visualized as a statistical spin system which consists of magnetic moments sitting at
fixed lattice points and interacting with their mean field, rather than as a gas of
itinerant electrons.

This picture is also the answer to a question that has often been raised, namely, how can
it be that there is a nontrivial RG flow in 3 spacetime dimensions even though the
gravitational field has no physical degrees of freedom in $d=3\,$? We explained this fact
by noting that a ``degree of freedom'' in the sense of this question amounts to a
\emph{propagating} degree of freedom or, in the language of the classical Cauchy problem,
to an independent component of the metric whose time development can be computed from the
field equations. While diamagnetic effects are in fact related to propagation and orbital
motion, and hence are indeed absent in $d=3$, the paramagnetic ones are still present in
that case. Since the interactions responsible for the nontrivial RG flow are,
essentially, only of paramagnetic nature, we now understand that they have nothing to do
with propagation, and so they do not count as degrees of freedom in the sense of the
question. They rather relate to the ultralocal spin orientation of the fluctuations
$h_\mn$ relative to a given background.

Another dimensionality of special interest is $d=2+\epsilon$. For $\epsilon\rightarrow 0$
Newton's constant is dimensionless and so one expects the leading order of $\eta_N$ to
become universal. Nevertheless, in the literature there has been a longstanding puzzle
about the correct $\ord(g)$ coefficient appearing in $\eta_N = -bg + \ord(g^2)$. In this
paper we have seen that actually both values for $b$ found by the majority of the
authors, namely $b = \frac{38}{3}$ and $b = \frac{2}{3}$, are correct in a certain sense.
However, they refer to two different running coupling constants, both of which make their
appearance in the effective average action: the coefficient $b = \frac{38}{3}$ belongs to
the bulk Newton constant $G_N$, while $b = \frac{2}{3} \equiv b^\p$ plays a similar
r\^{o}le for the boundary Newton constant $G_N^\p$ which occurs in the prefactor of the
Gibbons-Hawking term \cite{boundary}. The difference of $b$ and $b^\p$ is explained by
the fact that $b$ receives contributions from both the dia- and the paramagnetic
interactions, while $b^\p$ is nonzero due to the diamagnetic term alone.

Since, in $d < 3$ dimensions, the dia- and paramagnetic interactions drive $\eta_N$ in
the same direction, both of them contribute positively to $b$ in $d = 2+\epsilon$. This
observation makes it clear that the ``old'' fixed point with $b = \frac{2}{3}$ of
$2+\epsilon$ dimensional gravity found already in the 1970's is of a rather different
nature than the NGFP in 4 dimensions: the former exists only \emph{thanks to} the
diamagnetic interaction, the latter \emph{despite} it! Therefore, the fixed point with
$b=\frac{38}{3}$ can be seen as the dimensional continuation of the NGFP from
$d=4$ to $d=2+\epsilon$, while the one corresponding to $b=\frac{2}{3}$ can not.



\begin{thebibliography}{99}
%
\bibitem{Kiefer}
C. Kiefer, {\it Quantum Gravity}, Third Edition, Oxford Science Publications, Oxford (2012).
%
\bibitem{Hamber-book}
H.~W.~Hamber, \textit{Quantum Gravitation}, Springer, Berlin (2009).
%
\bibitem{A}
A.~Ashtekar, \textit{Lectures on non-perturbative canonical gravity,}
World Scientific,\\ Singapore (1991);\\
A.~Ashtekar and J.~Lewandowski, Class.\ Quant.\ Grav. 21 (2004) R53.
%
\bibitem{R}
C.~Rovelli, 
{\it Quantum Gravity,} Cambridge University Press, Cambridge (2004).
%
\bibitem{T}
Th.~Thiemann, 
{\it Modern Canonical Quantum General Relativity,}\\ Cambridge University Press, Cambridge (2007).
%
\bibitem{CDT}
J.~Ambj{\o}rn, J.~Jurkiewicz and R.~Loll, Phys. Rev. Lett. 93 (2004) 131301;
Phys. Lett. B 607 (2005) 205; Phys. Rev. Lett. 95 (2005) 171301;
Phys. Rev. D 72 (2005) 064014; Lect. Notes Phys. 807 (2010) 59;\\
J.~Ambj{\o}rn, S.~Jordan, J.~Jurkiewicz and R.~Loll, Phys. Rev. Lett. 107 (2011) 211303;\\
D.~Benedetti and J.~Henson, Phys. Rev. D 80 (2009) 124036.
%
\bibitem{Regge}
T.~Regge and R.~M.~Williams, J.\ Math.\ Phys.\ 41 (2000) 3964.\\
H.~W.~Hamber, Gen. Rel. Grav. 41 (2009) 817; Phys. Rev. D 45 (1992);
Phys. Rev. D 61 (2000) 124008; arXiv:0704.2895.
%
\bibitem{Weinberg}
S.~Weinberg 
in \textit{General Relativity, an Einstein Centenary Survey},
S.~W.~Hawking and W.~Israel (Eds.), Cambridge University Press (1979). 
%
\bibitem{weinberg-book}
S.~Weinberg, \textit{Gravitation and Cosmology}, Wiley, New York (1972).
%
\bibitem{mr}
M.~Reuter, Phys.\ Rev.\ D 57 (1998) 971 and \mbox{hep-th/9605030}.
%
\bibitem{elisa1}
  E.~Manrique and M.~Reuter, Phys.\ Rev.\  D 79 (2009) 025008 and \mbox{arXiv:0811.3888}.
%
\bibitem{souma}
W.~Souma,
Prog.\ Theor.\ Phys.\ 102 (1999) 181.
%
\bibitem{frank1}
M.~Reuter and F.~Saueressig, 
Phys.\ Rev.\ D 65 (2002) 065016 and 
\mbox{hep-th/0110054.}
%
\bibitem{oliver}
O.~Lauscher and M.~Reuter, Phys.\ Rev.\ D 65 (2002) 025013, \mbox{hep-th/0108040};
Phys. Rev. D 66 (2002) 025026, \mbox{hep-th/0205062};
Class. Quant. Grav. 19 (2002) 483, \mbox{hep-th/0110021}.
%
\bibitem{NJP}
For a recent review on QEG and Asymptotic Safety with a comprehensive reference list see\\
M.~Reuter and F.~Saueressig, New J. Phys. 14 (2012) 055022 and arXiv:1202.2274.
%
\bibitem{livrev}
M.~Niedermaier and M.~Reuter, Living Reviews in Relativity 9 (2006) 5.
%
\bibitem{reviews}
For further reviews on Asymptotic Safety see:\\
M.~Reuter and F.~Saueressig, in \textit{Geometric and Topological Methods for Quantum
Field Theory}, H.~Ocampo, S.~Paycha and A.~Vargas (Eds.), Cambridge Univ. Press,
Cambridge (2010), arXiv:0708.1317;\\
R.~Percacci, in \textit{Approaches to Quantum Gravity: Towards a New Understanding of
Space, Time and Matter}, D.~Oriti (Ed.), Cambridge Univ. Press, Cambridge, 2009,
arXiv:0709.3851;
%
\bibitem{CPR}
A.~Codello, R.~Percacci and C.~Rahmede, Ann. Phys. 324 (2009) 414.
%
\bibitem{elisa2}
E.~Manrique and M.~Reuter, Annals Phys. 325 (2010) 785 and \mbox{arXiv:0907.2617}.
%
\bibitem{MRS1}
E.~Manrique, M.~Reuter and F.~Saueressig, Annals Phys. 326 (2011) 440, \mbox{arXiv:1003.5129}.
%
\bibitem{MRS2}
E.~Manrique, M.~Reuter and F.~Saueressig, Annals Phys. 326 (2011) 463, \mbox{arXiv:1006.0099}.
%
\bibitem{dewitt-books}
B.~S.~DeWitt, \textit{The Global Approach to Quantum Field Theory}, Clarendon Press, Oxford (2003).
%
\bibitem{Lawson}
H.~B.~Lawson and M.-L.~Michelsohn, \textit{Spin Geometry}, Princeton Univ.\ Press, Princeton (1989).
%
\bibitem{Nielsen}
N.~K.~Nielsen, Am. J. Phys. 49 (1981) 1171.
%
\bibitem{polyakov-book}
A.~M.~Polyakov, \textit{Gauge Fields and Strings}, Harwood, Chur (1987).
%
\bibitem{Johnson}
K.~Johnson, in \textit{Asymptotic Realms in Physics}, Eds.\ A.~Guth, K.~Huang and R.~L.~Jaffe,
MIT Press, Cambridge (1983).
%
\bibitem{huang-book}
K.~Huang, \textit{Quarks, Leptons and Gauge Fields}, World Scientific, Singapore (1992).
%
\bibitem{Gottfried-Weisskopf}
K.~Gottfried and V.~F.~Weisskopf, \textit{Concepts of Particle Physics: Volume II},
Oxford University Press (1986).
%
\bibitem{dittrich}
W.~Dittrich and M.~Reuter, Phys.~Lett.~B 128 (1983) 321.
%
\bibitem{mrtheta}
M.~Reuter, Mod. Phys. Lett. A12 (1997) 2777.
%
\bibitem{heat-kernel}
G.~A.~Vilkovisky, CERN preprint TH-6392-92 (1992).
%
\bibitem{PolitzerGrossWilczek}
H.~D.~Politzer, Phys. Rev. Lett. 30 (1973) 1346;\\
D.~Gross and F.~Wilczek, Phys. Rev. Lett. 30 (1973) 1343.
%
\bibitem{avact}
C.~Wetterich, Phys. Lett. B 301 (1993) 90-94.
%
\bibitem{YM-EAA}
M.~Reuter and C.~Wetterich, Nucl. Phys. B 417 (1994) 181;
Nucl. Phys. B 427 (1994) 291;
Phys. Rev. D 56 (1997) 7893.
%
\bibitem{YM-AA}
M.~Reuter and C.~Wetterich, Nucl. Phys. B 391 (1993) 147; Nucl. Phys. B 408 (1993) 91.
%
\bibitem{GiesJaeckel}
H.~Gies and J.~Jaeckel, Phys. Rev. Lett. 93 (2004) 110406.
%
\bibitem{GiesYM}
H.~Gies, Phys. Rev. D 68 (2003) 085015.
%
\bibitem{HE}
W.~Heisenberg and H.~Euler, Z. Phys. 98 (1936) 714.
%
\bibitem{opt}
D.~F.~Litim, Phys. Rev. D 64 (2001) 105007.
%
\bibitem{boundary}
D.~Becker and M.~Reuter, \mbox{arXiv:1205.3583}.
%
\bibitem{GKT}
R.~Gastmans, R.~Kallosh and C.~Truffin, Nucl. Phys. B 133 (1978) 417.
%
\bibitem{CD}
S.~M.~Christensen and M.~J.~Duff, Phys. Lett. B 79 (1978) 213.
%
\bibitem{Brown}
L.~Brown, Phys. Rev. D 15 (1977) 1469.
%
\bibitem{Tsao}
H.-S.~Tsao, Phys. Lett. B 68 (1977) 79.
%
\bibitem{Kawai}
H.~Kawai and M. Ninomiya, Nucl. Phys. B 336 (1990) 115.
%
\bibitem{JackJones}
I.~Jack and D.~R.~T.~Jones, Nucl. Phys. B 358 (1991) 695.
%
\bibitem{membrane}
A.~Codello and O.~Zanusso, Phys. Rev. D 83 (2011) 125021.
%
\bibitem{GH}
G.~W.~Gibbons and S.~W.~Hawking, Phys. Rev. D 15 (1977) 2752.
%
\bibitem{wet-dec}
C.~Wetterich, Z. Phys. C 57 (1993) 451.
%
\bibitem{h1}
M.~Reuter and H.~Weyer, Phys. Rev. D 69 (2004) 104022, hep-th/0311196;\\
Phys. Rev. D 70 (2004) 124028, hep-th/0311196.
%
\bibitem{DR1}
W.~Dittrich and M.~Reuter, \textit{Effective Lagrangians in Quantum Electrodynamics}, Springer, Berlin (1985).
%
\bibitem{bh}
A.~Bonanno and M.~Reuter,
Phys.\ Rev.\ D 62 (2000) 043008; Phys.\ Rev.\ D 73 (2006) 083005;
Phys.\ Rev.\ D 60 (1999) 084011.
%
\bibitem{erick}
M.~Reuter and E.~Tuiran, Phys.\ Rev.\ D 83 (2011) 044041, arXiv:1009.3528;\\
in \emph{Proceedings of the Eleventh Marcel Grossmann Meeting}, H.~Kleinert, R.~Jantzen,
R.~Ruffini (Eds.), World Scientific, Singapore (2007), hep-th/0612037.
%
\bibitem{don}
J.~F.~Donoghue, Phys. Rev. Lett. 72 (1994) 2996; Phys. Rev. D 50 (1994) 3874.
%
\bibitem{alfio}
A.~Bonanno, Phys. Rev. D 85 (2012) 081503.
%
\bibitem{Lor-EAA}
E.~Manrique, S.~Rechenberger and F.~Saueressig, Phys. Rev. Lett. 106 (2011) 251302.
%
\end{thebibliography}
\end{document}